\documentclass[iop]{emulateapj}

\usepackage{graphicx}
\usepackage{amsmath}
\usepackage{amssymb}
\usepackage{latexsym}
\usepackage{epsfig}
\usepackage{natbib}

\newcommand{\half}{\frac{1}{2}}

\newcommand{\xv}{\mathbf{x}}

\newcommand{\lv}{\boldsymbol{\ell}}
\newcommand{\vvec}{\boldsymbol{v}}
\newcommand{\uvec}{\boldsymbol{v}}
\newcommand{\kvec}{\boldsymbol{k}}
\newcommand{\rvec}{\boldsymbol{r}}

\shorttitle{}
\shortauthors{F. Miniati}

\begin{document}
\title{The Matryoshka Run (II): Time Dependent Turbulence Statistics, Stochastic Particle Acceleration and
Microphysics Impact in a Massive Galaxy Cluster}

\author{Francesco Miniati}
\affil{Physics Department, Wolfgang-Pauli-Strasse 27,
ETH-Z\"urich, CH-8093, Z\"urich, Switzerland; fm@phys.ethz.ch}

\begin{abstract}
We use the Matryoshka run to study the time dependent statistics of structure-formation driven 
turbulence in the intracluster medium of a 10$^{15}M_\odot$ galaxy cluster.
We investigate the turbulent cascade in the inner Mpc 
for both compressional and incompressible velocity components. 
The flow maintains approximate conditions of fully developed turbulence, with
departures thereof settling in about an eddy-turnover-time.
Turbulent velocity dispersion remains above $700$ km s$^{-1}$ even at low mass 
accretion rate, with the fraction of compressional energy between 10\% and 40\%.
Normalisation and slope of compressional turbulence is susceptible to large variations on short time scales, unlike the incompressible counterpart.
A major merger occurs around redshift $z\simeq0$ and is
accompanied by a long period of enhanced turbulence, ascribed to temporal 
clustering of mass accretion related to spatial clustering of matter. We test models of 
stochastic acceleration by compressional modes for the origin of diffuse radio emission in galaxy clusters.
The turbulence simulation model constrains an important unknown of this complex problem
and brings forth its dependence on the elusive micro-physics of the intracluster plasma.
In particular, the specifics of the plasma collisionality and the dissipation physics of weak shocks
affect the cascade of compressional modes with strong impact on the acceleration rates.
In this context radio halos emerge as complex phenomena in which a hierarchy of processes acting on
progressively smaller scales are at work.
Stochastic acceleration by compressional modes
implies statistical correlation of radio power and spectral index 
with merging cores distance, both testable in principle with radio surveys.

\end{abstract}

\section{Introduction} \label{intro:sec} 

Astrophysical flows are well known to be turbulent for the most part.
However, with due exceptions~\citep{Porter00,Viallet13,Bodo11,Parkin13}, 
this character of the flow is generally not reproduced in numerical
models. The reason is simply due to the impractical computational cost. Turbulence 
requires the resolution of a large range of spacial scales, because it does
not develop unless the flow's Reynolds number {\it Re} is sufficiently large.
Nevertheless, there has been growing attention to the role of turbulence in recent years,
calling for renewed efforts in achieving high {\it Re} flows in numerical models.
This is particularly the case of the intra-cluster medium (ICM),
the hot and magnetised plasma pervading the volume of galaxy clusters (GC).

In the ICM, as in other like flows,
turbulence regulates the magnetic field energy density and structure~\citep{2006MNRAS.366.1437S,IapichinoNiemeyer2008}. 
Knowledge of the statistical properties of the turbulence and of its time evolution is 
essential to evaluate fundamental quantities such as the equipartition scale between 
magnetic and kinetic energy~\citep{2008Sci...320..909R,2009ApJ...693.1449C,2012MNRAS.422.3495B}.
Also, turbulence acts as an effective non-thermal pressure supporting the ICM
gas against self-gravity~\citep{Faltenbacher05,Dolag05b,Nagai07,Lau09,Biffi11,Valdarnini11}. 
This induces a bias in the mass estimates of GC which propagates as a
systematic error when such estimates are employed in cosmology.
Again, proper resolution of the turbulent flow is important to estimate the turbulent energy density
across the relevant scales and accurately determine the above biases~\citep{2014ApJ...782...21M}.
Finally, cluster turbulence influences the transport of relativistic particles both in coordinate
and momentum space.
In fact, a significant fraction of GC betray the presence of giant halos of diffuse radio emission
extending over distances $\sim$Mpc~\citep{2008SSRv..134...93F,2012A&ARv..20...54F,2014IJMPD..2330007B}. The nature of this phenomenon is 
unknown but, upon inspection, it reveals that relativistic electrons must be accelerated in the 
ICM~\citep{1977ApJ...212....1J}.  In addition, the phenomenon appears to be triggered 
by major merger events~\citep[][see also~\cite{2011MNRAS.417L...1R,2012MNRAS.421L.112B}]{2001ApJ...553L..15B,2003A&A...399..813F,2006AJ....131.2900C,2010ApJ...721L..82C}. Amongst other proposals~\citep{1999APh....12..169B,2001ApJ...562..233M,2010ApJ...722..737K,2008MNRAS.385.1211P,2011A&A...527A..99E} a compelling and popular
idea for the origin of radio halos (RH) is that the relativistic particles responsible for the radio emission 
are accelerated by turbulence~\citep[e.g.][]{1987A&A...182...21S} particularly when 
generated in connection with a major merger event~\citep[e.g.][]{2001MNRAS.320..365B,2001ApJ...557..560P}. In particular, in recent years, much effort has been devoted to models based on
compressional turbulence~\cite[][referred to in the following as BL7 and BL11, respectively]{2007MNRAS.378..245B,2011MNRAS.410..127B}.

Several numerical studies have already addressed such connection between merger driven
turbulence and the origin of RH~\cite[e.g.][]{2009A&A...504...33V,2011ApJ...726...17P,2011A&A...529A..17V,
Hallman11,2013ApJ...771..131B,2013MNRAS.429.3564D,2014MNRAS.443.3564D}.
In particular~\cite{2013ApJ...771..131B} employed a sophisticated spectral decomposition
scheme to estimate the acceleration rates from compressible modes using results from 
an MHD-AMR cosmological simulation of a galaxy cluster. In their analysis they focus on an 
individual time frame, hence the time evolution is not explored,
and study in detail the role of the cosmic-ray feedback on the turbulence.
In addition,~\cite{2013MNRAS.429.3564D} and~\cite{2013ApJ...762...78Z} have carefully
followed the stochastic acceleration of relativistic particles during merger and sloshing motions,
respectively, using more approximate turbulence, mode decomposition and cluster models in their
non-comological simulations.
A numerical model is useful in that it provides an estimate of the turbulent energy
in the ICM. This quantity impacts the transport coefficients, hence the acceleration rates of particle. 
In the aforementioned acceleration models, it is the compressional component of
the turbulent motion that interacts with, and supplies
energy to, the relativistic particles. In addition, the calculation of the acceleration rates requires
knowledge of other characteristics of the turbulence cascade, namely the outer scale, velocity
dispersion and slope. 
In order to determine this quantities, the numerical model must resolve the turbulence cascade.
In addition, these quantities should ideally be determined as a function of time during the GC's 
formation history.
It is, in fact, the rise of turbulent energy, following a major merger that is believed to power RH.
Thus it is important to determine how the turbulent cascade  
changes in normalisation and slope in response to the changing conditions of 
mass accretion rate and merger events. 
This important aspect of the problem remains to be studied and is the objective of this paper. 

Indeed, thanks to advances in computational techniques and supercomputer power we have 
recently been able to model the statistical properties of ICM turbulence in a fully cosmological 
context~\citep{2014ApJ...782...21M}. The focus of the calculation is the formation 
and evolution of a massive GC with virial mass, $M=1.3\times 10^{15}M_\odot$ at redshift, $z=0$. 
This system is particularly interesting from the
viewpoint of the origin of RH, because it undergoes a major merger around $z=0$. 
In the first paper~\citep{2014ApJ...782...21M} we have introduced the methodology adopted 
in carrying out the numerical calculation, and presented the statistical analysis of the ICM 
turbulence for a single time snapshot corresponding to $z\simeq 0$.

In this paper we study, instead, the time evolution of the statistical properties of the ICM turbulence.
In doing so we give priority to those properties most relevant to the acceleration of particles. In fact,
in the second part of the paper we use our findings on the ICM turbulence to explore their implications
in terms of particle acceleration and, possibly, on the origin of RH. We explore in particular two 
acceleration models, the non-resonant mechanism~\cite[NR,][]{1988SvAL...14..255P} and 
transient-time-damping~\cite[TTD,][]{1976JGR....81.4633F,1991ApJ...376..342M}, 
both advocated and studied in great details for the ICM case in BL7 and BL11.
As already anticipated, both processes are powered by the compressional component of the turbulence.
However, additional physics of the ICM plasma is at work here, which is not completely 
established and which is not directly modelled by the numerical simulation.
We have therefore explored different assumptions about such physics and found that they
have important consequences on the results.
In fact, one of the great advantages and most valuable contribution of a high resolution numerical model
such as this, is that  by resolving the turbulence cascade of compressional and 
incompressible motions it narrows an important unknown of this complex problem, 
and brings forth the implications of other assumptions concerning the 
elusive and difficult to test micro-physics of the ICM plasma.

The reminder of this paper is organised as follows.
In Section~\ref{method:sec} we introduce the methods employed in
the paper, including the details of the numerical simulation (\ref{numerics:sec}) and the
analysis to extract the turbulent statistics (\ref{analysis:sec}). In Section~\ref{res:sec} we present the 
results concerning the time-evolution of the GC: this includes the mass accretion history,
the evolution of the turbulent kinetic energy and the fraction in compressional motions thereof
(\ref{ah:sec}); as well as the evolution of the turbulence statistics such as second order
structure functions, slope of the cascade inertial range, pdf of the velocity increments at 
the turbulence outer scale (\ref{tds:sec}). Section~\ref{resII:sec} presents the 
second part of results, those concerning the acceleration of particles. 
In particular we introduce the diffusion coefficients
(\ref{dpp:sec}) and cascade models (\ref{camo:sec}) used in estimating the particle advection
rates in momentum space and discussing their implications in connection with the origin of RH
(\ref{adr:sec}). We discuss the results in Section~\ref{t_disc:sec} and conclude with a summary
and conclusions in Section~\ref{sum:sec}.

\section{Methods} \label{method:sec}
\subsection{Numerics} \label{numerics:sec}

The Matryoshka run, including the novel Eulerian refinement approach,
is fully described in~\cite{2014ApJ...782...21M}
The run is carried out with~\texttt{CHARM}
an Adaptive-Mesh-Refinement code for cosmology,
based on a directionally unsplit variant of the Piecewise-Parabolic-Method for
hydrodynamics~\citep{1990JCoPh..87..171C}, a time centered modified symplectic
scheme for the collisionless dark matter and solve Poisson's equation
with a second order accurate discretization~\citep{2007JCoPh.227..400M}. 
We also evolve a dynamically negligible magnetic field using the
constrained-transport algorithm for solenoidal MHD described
in~\cite{2011ApJS..195....5M}. Radiative cooling and heating are neglected.
The cosmological model corresponds to 
a concordance $\Lambda$-CDM universe with normalized (in
units of the critical value) total mass density, $\Omega_m=0.2792$,
baryonic mass density, $\Omega_b=0.0462$, vacuum energy density,
$\Omega_\Lambda= 1- \Omega_m= 0.7208$, normalized Hubble constant
$h\equiv H_0/100$ km s$^{-1}$ Mpc$^{-1}$ = 0.701, spectral index of
primordial perturbation, $n_s=0.96$, and rms linear density
fluctuation within a sphere with a comoving radius of 8 $\,h^{-1}$
Mpc, $\sigma_8=0.817$~\citep{Komatsu09}.
The computational volume comoving size is 
$L_{Box}=240\,h^{-1}$ Mpc. Here the baryonic and dark matter variables are
initialised with~\texttt{grafic++} (a parallel version of \texttt{grafic2}
\citep{Bertschinger01}, made publicly available by D. Potter)
using the power spectrum interpolation method suggested in~\cite{EisensteinHu98}.

The purpose of the simulation is to model at high resolution the formation of a massive GC.
The latter was identified in a preliminary low resolution run
at redshift zero with our implementation of the HOP halo finder~\citep{1998ApJ...498..137E}
Thus we generate zoom-in type of initial conditions,
on three progressively finer and uniform resolution grids. The coarsest 
level grid is made of 512$^3$ comoving cells, corresponding to a nominal spatial resolution
of 468.75$\,h^{-1}$ comoving kpc, and the collisionless dark matter component is represented 
with 512$^3$ particles of mass $6.7\times 10^9\,h^{-1}$ M$_\odot$. 
A region of space surrounding the Lagrangian volume of the target GC, is then 
initialised at higher resolution with the help of two additional levels of refinement 
on top of the base grid. 
The refinement ratio for both levels is, $n_\mathrm{ref}^\ell\equiv \Delta x_{\ell}/\Delta x_{\ell+1}=2$, $\ell=0,1$.
Each refined level covers 1/8 of the volume of the next coarser level
with a uniform grid of 512$^3$ comoving cells while the dark matter
distribution function is represented by 512$^3$ particles.
At the finest level the spatial resolution is $\Delta x=$ 117.2$\,h^{-1}$ comoving kpc
and the particle mass is $1.0\times 10^8\,h^{-1}$ M$_\odot$.
In addition, as the Lagrangian volume of the GC shrinks under self-gravity, three additional 
uniform grids covering 1/8 of the volume of the next coarser level
with 512$^3$, 1024$^3$ and 1024$^3$ comoving cells are employed,
with a corresponding refinement ratio n$_\mathrm{ref}^\ell=2,4,2$, for $\ell=2,3,4$, respectively.
All of these additional three grids are in place by redshift 1.4, providing a spatial resolution 
of 7.3 h$^{-1}$ comoving kpc in a volume of 7.5 h$^{-1}$ Mpc, which accommodates the whole 
virial volume of the GC. The ensuing dynamic range of resolved spatial scales is sufficiently large for the 
emergence of  turbulence.

\subsection{Analysis of Turbulence Velocity Field}\label{analysis:sec}

The exquisite resolution across the GC volume allows us to accurately measure the time dependent 
statistical properties of the turbulent motions. In particular we 
compute the longitudinal and transverse second order structure functions for both 
the solenoidal and compressional components of the velocity. To do so we first carry out the
following Hodge-Helmholtz decompositions,
\begin{gather} \label{hh0:eq}
\vvec=\vvec_s+\vvec_c, \\
\vvec_c=-\nabla\phi, \quad \vvec_s=\nabla\times \mathbf{A}, \\
\phi =\frac{1}{4\pi}\int \frac{\nabla\cdot\vvec}{r} d\xv, \quad
\mathbf{A} =\frac{1}{4\pi}\int \frac{\nabla\times\vvec}{r} d\xv. \label{hh2:eq}
\end{gather}
Then, for each component we can define the second order rank correlation function 
of velocity increments~\citep{LandauLifshitz6}
\begin{equation}\label{gcf:eq}
S_{ij}(\xv,\lv) \equiv \left\langle \delta\vvec _i \delta\vvec _j \right\rangle,
\end{equation}
where ${i}$ indicates the vector component along $i$-th axis and
$\delta\uvec_i\equiv [\uvec(\xv+\lv)-\uvec(\xv)]_i$.
For a homogeneous and isotropic flow this tensor is only a
function of $\ell=|\lv|$ and can be expressed in terms of the
longitudinal and transverse correlation functions~\citep{LandauLifshitz6},
\begin{equation}\label{gicf:eq}
S_{ij}(\ell)=S^{(2)}_{l}(\ell)n_in_j+S^{(2)}_{t}(\delta_{ij}-n_in_j)
\end{equation}
where $S^{(2)}_{l}$ and $S^{(2)}_{t}$ are computed by taking in Eq. \ref{gcf:eq} 
the velocity components parallel and perpendicular to $\lv$, respectively.
For isotropic and homogeneous turbulence one can also write~\citep{LandauLifshitz6}
\begin{equation}\label{lsf2:eq}
S_{ij}(\ell)=\frac{2}{3}\langle (\delta\uvec)^2 \rangle \delta_{ij}
-2 s_{ij}(\ell)
\end{equation}
where $s_{ij}(\ell)=\left\langle [\uvec(\xv)]_i  [\uvec(\xv+\lv)]_j\right\rangle.$ 
Since the velocity components become uncorrelated at large distances, the asymptotic behaviour is as follows
\begin{gather}
\lim_{\ell\to \infty}{s_{ij}=0},\quad \lim_{\ell\to \infty}{S_{ij}=\frac{2}{3}\langle(\delta\uvec)^2 \rangle\delta_{ij}}.
\end{gather}
The asymptotic behaviour of $S_{ij}$ can be reliably computed from the numerical simulations, which 
allows to also accurately infer $s_{ij}$. The total second order velocity correlation function is then given 
by
\begin{equation}\label{tsf:eq}
s^{(2)}(\ell)=s_\ell^{(2)}(\ell) + 2s_t^{(2)}(\ell).
\end{equation}
The correlation functions in configuration space can be converted into a power spectrum of isotropic
velocity fluctuations through a Fourier transformation, namely,
\begin{gather}
s^{(2)}(k)=\int_0^{\infty}d^3r \,s^{(2)}(r)\,e^{-i\kvec\cdot\rvec}= s_0 \left(\frac{k}{k_L}\right)^{-\zeta_2-3},
\end{gather}
where the last equality applies to the inertial range, $k_L$ is the mode corresponding to the largest scale 
and $\zeta_2$, the slope of the second order structure function. The integral can be carried through 
to determine $s_0$ from the normalisation of $s(\ell)$
\begin{gather}\label{ke:eq}
\langle(\delta\uvec)^2 \rangle = \int dk\,4\pi\, k^2\,s^{(2)}(k),
\end{gather}
so that
\begin{gather}
s_0=\frac{\zeta_2\langle(\delta\uvec)^2 \rangle}{4\pi k_L^3}.
\end{gather} 
Note that with the above notation the kinetic energy spectrum is
\begin{equation}\label{ek:eq}
E(k)=\zeta_2\frac{\langle(\delta\uvec)^2\rangle}{2k_L} \left(\frac{k}{k_L}\right)^{-\zeta_2-1}.
\end{equation}
In addition, in the following certain quantities are computed in terms of the total energy spectrum, $W(k)$,
which includes in addition to the kinetic component also the potential component associated to pressure 
fluctuations. To good approximation the two are in equipartition~\cite[e.g.][]{1991JFM...227..473S},
which will be assumed here (see also BL7), i.e.
\begin{equation}\label{wek:eq}
W(k)\approx 2 E(k).
\end{equation}
To compute the structure function we define sampling points randomly distributed inside the volume of 
interest and compute the velocity difference with respect to randomly selected field points at a maximum 
distances of two virial radii. A total of 10$^6$ sampling points and 10$^6$ field points are used for this 
purpose for a total of 10$^{12}$ pairwise velocity increments.
In our analysis we do not attempt to filter out large scale uniform motions 
explicitly~\cite[see, e.g.][]{2011A&A...529A..17V}. 
The structure functions are actually designed to do that and $S(\ell)$
does not include contributions from flows that are coherent above the scale, $\ell$. 
A comparison with other methods in the literature is beyond the scope of the paper.
However, it is worth noting that our results, presented in Section~\ref{tds:sec}, indicate 
that the scales at which the structure functions flatten, which we identify with the turbulence
outer scale, are consistent
with the results in, e.g.,~\cite{2011A&A...529A..17V}, who use a low-pass filter  (the k-max method)
to remove what they considered large scale motions unrelated to turbulence.
In the following we restrict the analysis of the velocity field, including the 
velocity structure functions, to the region within 1 Mpc from the GC centre. 
This is the most interesting region from the point of view of astronomical observations.
In particular, the diffuse radio-halo emission is observed primarily within this limited volume
and in the final part of the paper we attempt to apply our findings concerning the properties
of turbulence to models of particle acceleration by compressional motions.

Note, that the statistics of turbulence can also be charactersed in terms of 
power spectra obtained through an analysis in Fourier space. This analysis
is in principle equivalent since power spectra and structure functions are 
related through a Fourier transformation. Fourier analysis is less expensive
(scaling with the number of points $N$ as $N\times \log(N)$),
hence in principle preferable, but it requires periodic boundary conditions
which in the case of a cluster have to be constructed artificially.
Structure functions are more expensive (scaling instead as N$^2$)
but do not require a choice of boundary conditions, proving particularly
useful to compute the turbulence statistics in a selected region of
the galaxy cluster volume (like in a thick shell).

\begin{figure}[t] 
\centering
\includegraphics[width=0.475\textwidth,angle=0]{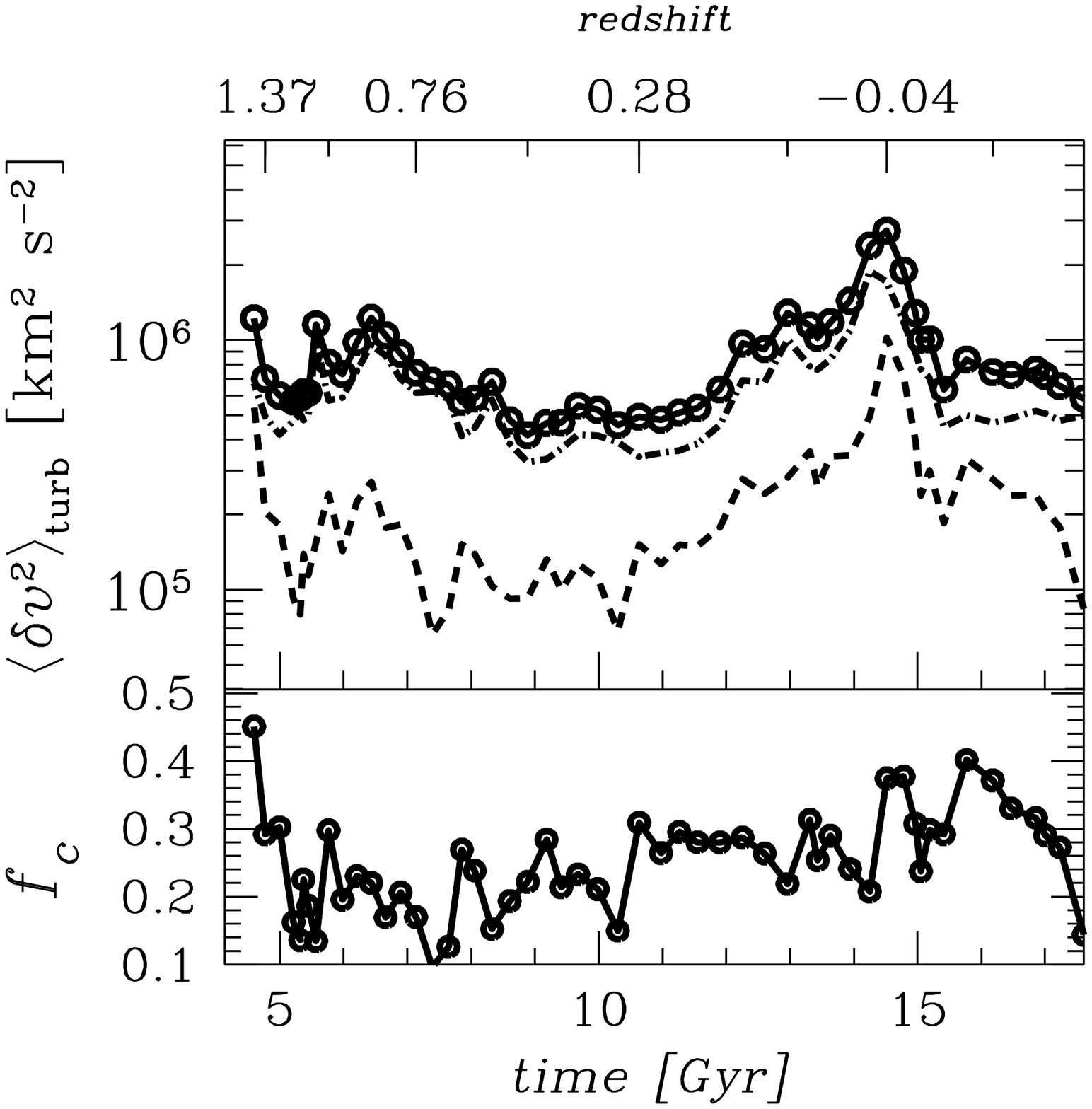}
\caption{
{\it Top:} Time evolution of the turbulent velocity dispersion. The incompressible and compressional
components are represented with a dot-dashed and dashed line, respectively, and their sum with 
a solid line and open dot symbols.
{\it Bottom:} Fraction of turbulent velocity dispersion in the compressional component.
\label{f1:fig}}
\end{figure}
\begin{figure*}[t] 
\centering
\includegraphics[width=0.75\textwidth,angle=0]{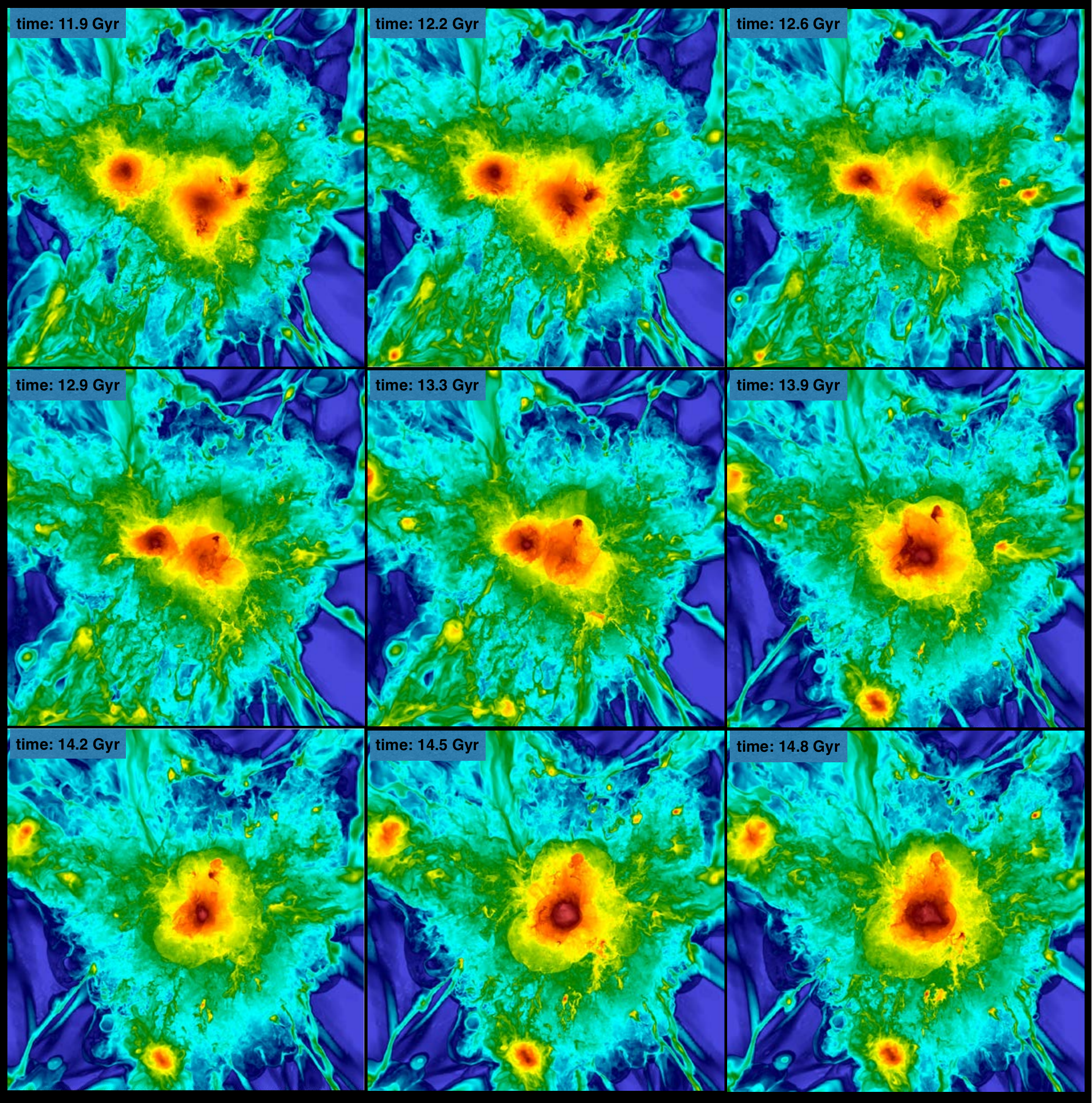}
\caption{
Merger history during the redshift interval $0.15 \lesssim z\lesssim -0.06$.
Each panel shows the density distribution in log scale on a plane passing 
approximately through the GC centre. Time (in the top-left of each panel) increases from top-left to 
bottom-right. The top-left panel shows the main GC 
at the centre, a large substructure on the left with mass ration 1:2.5 and two smaller substructures: 
the largest with mass ratio 1:10 to the top-right (SubA) of the main GC and the other (SubB) below it. 
The first merger experienced by the GC is due to the accretion of these two structures and
culminates at t=12.9 billion yr (middle-left panel). The the second merger with the largest substructure 
is in full swing by t=13.9 billion yr (middle right), with core passage around t=14.2 billion yr (bottom left).
\label{f2:fig}}
\end{figure*}
\begin{figure*}[t]
\centering
\includegraphics[width=0.5\textwidth]{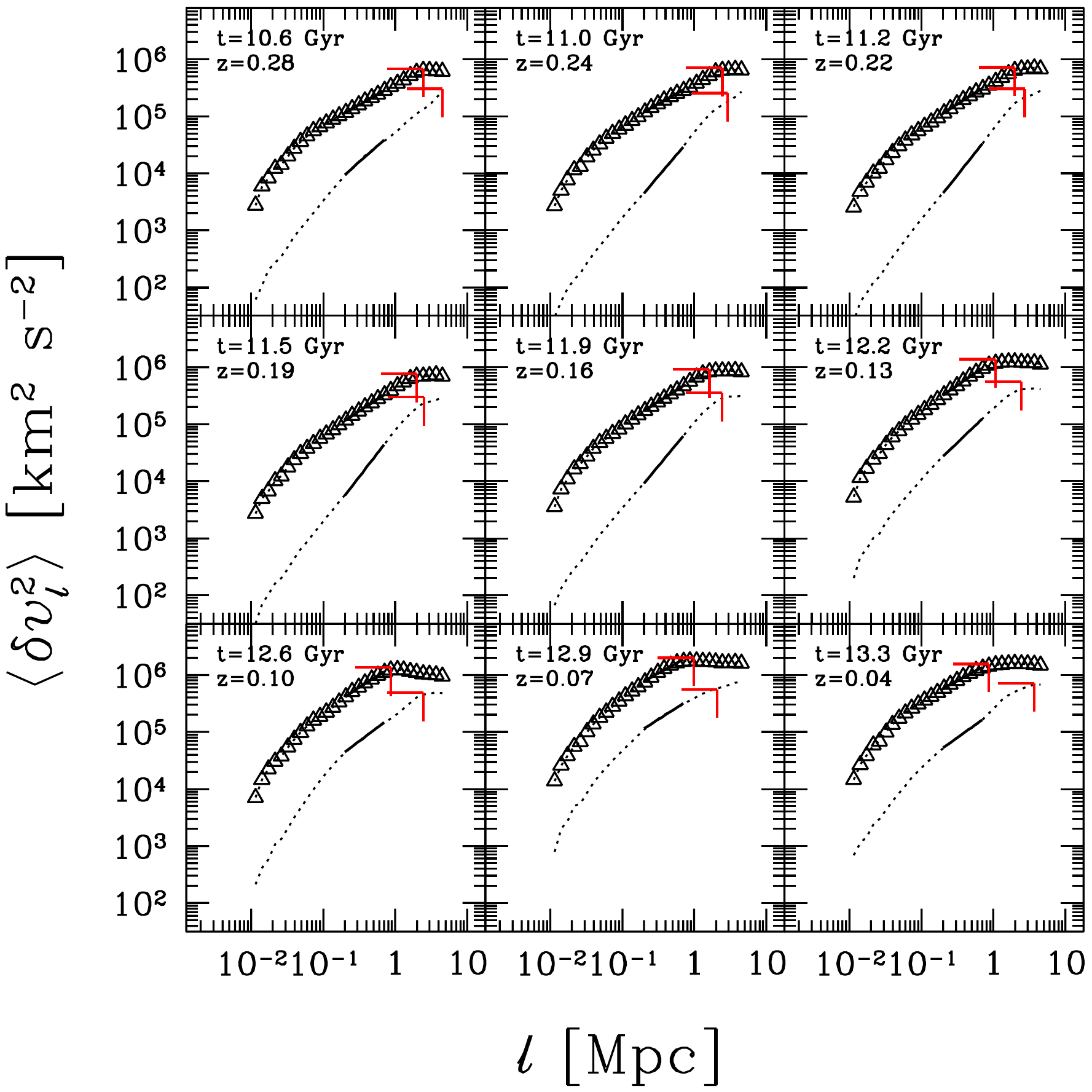}\includegraphics[width=0.5\textwidth]{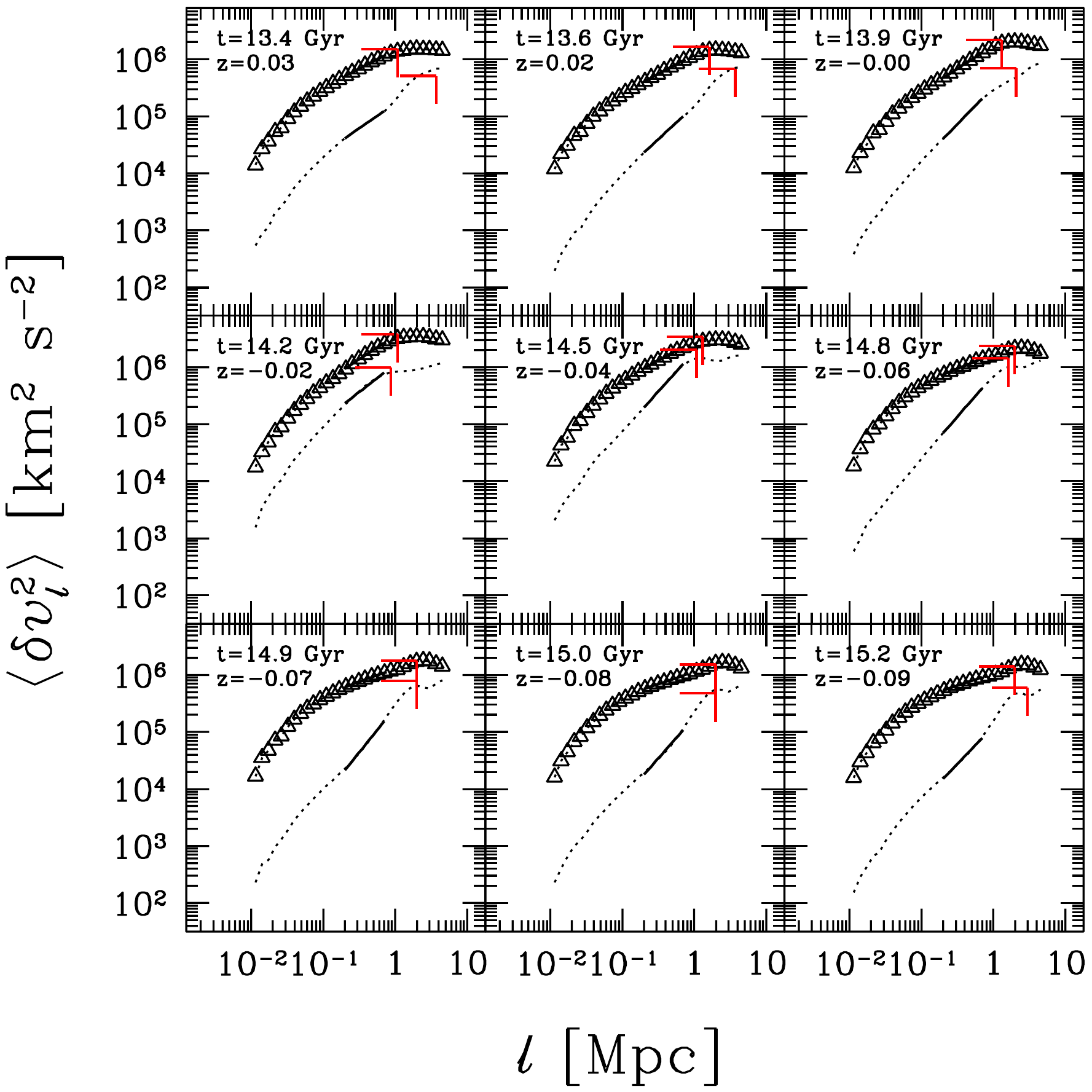}
\caption{
Total second order velocity structure functions (see Eq.~\ref{tsf:eq})
for solenoidal (triangles) and compressional (dots) components, respectively. 
For each component, the solid line represents the normalisation and slope in the inertial range extracted
by our fitting procedure, while the vertical and horizontal red line the injection scale and the corresponding velocity increment. Time and redshift are indicated in the top left corner of each panel.
\label{f3:fig} }
\end{figure*}
\begin{figure}[t]
\centering
\includegraphics[width=0.475\textwidth]{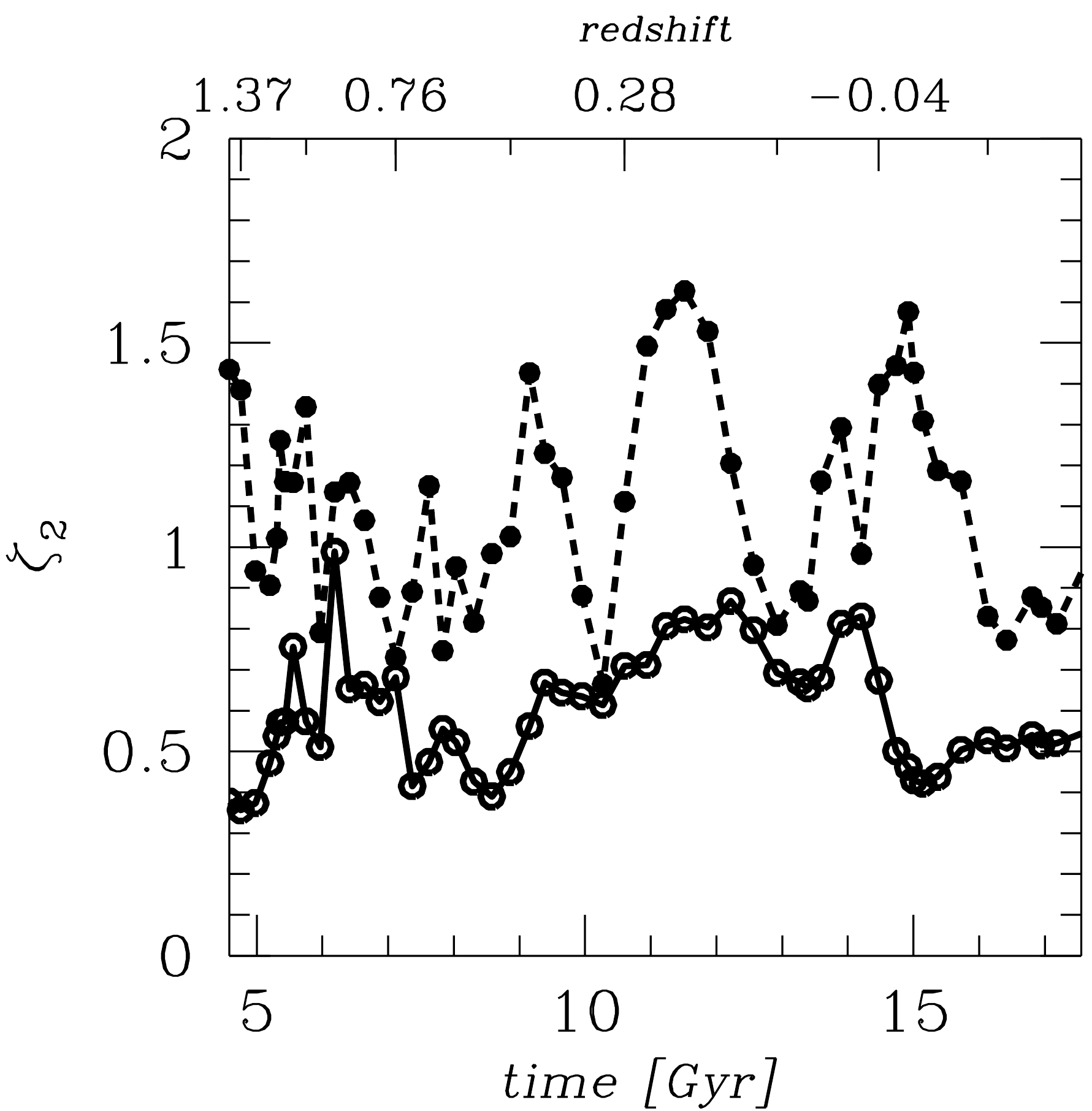}
\caption{
Slope of the structure functions' inertial range as a function of time for both the incompressible
(solid line and open symbols) and compressional (dashed line and solid symbol) velocity
components.
\label{f4:fig}
}
\end{figure}
\begin{figure*}[t]
\centering
\includegraphics[width=0.5\textwidth]{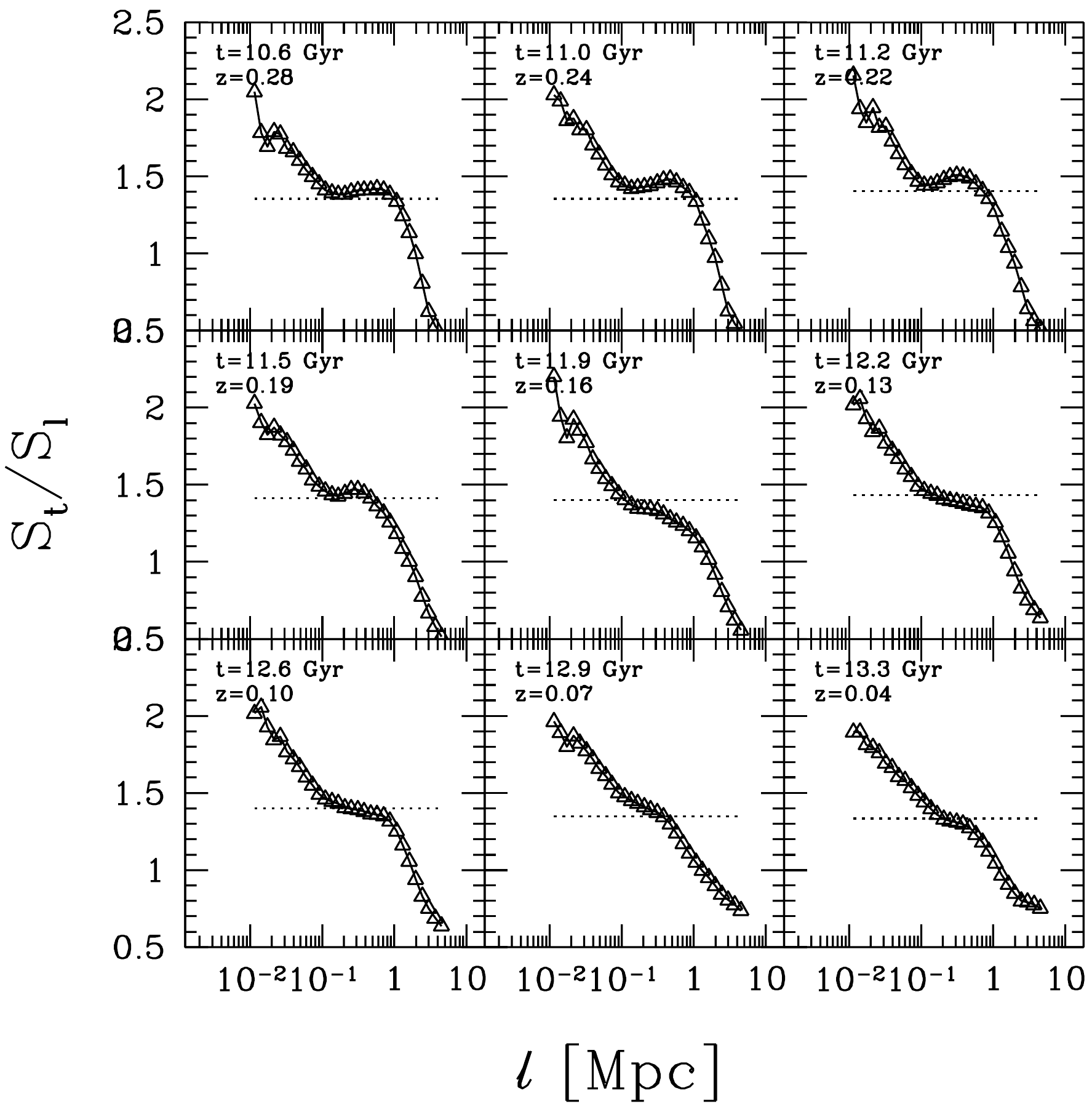}\includegraphics[width=0.5\textwidth]{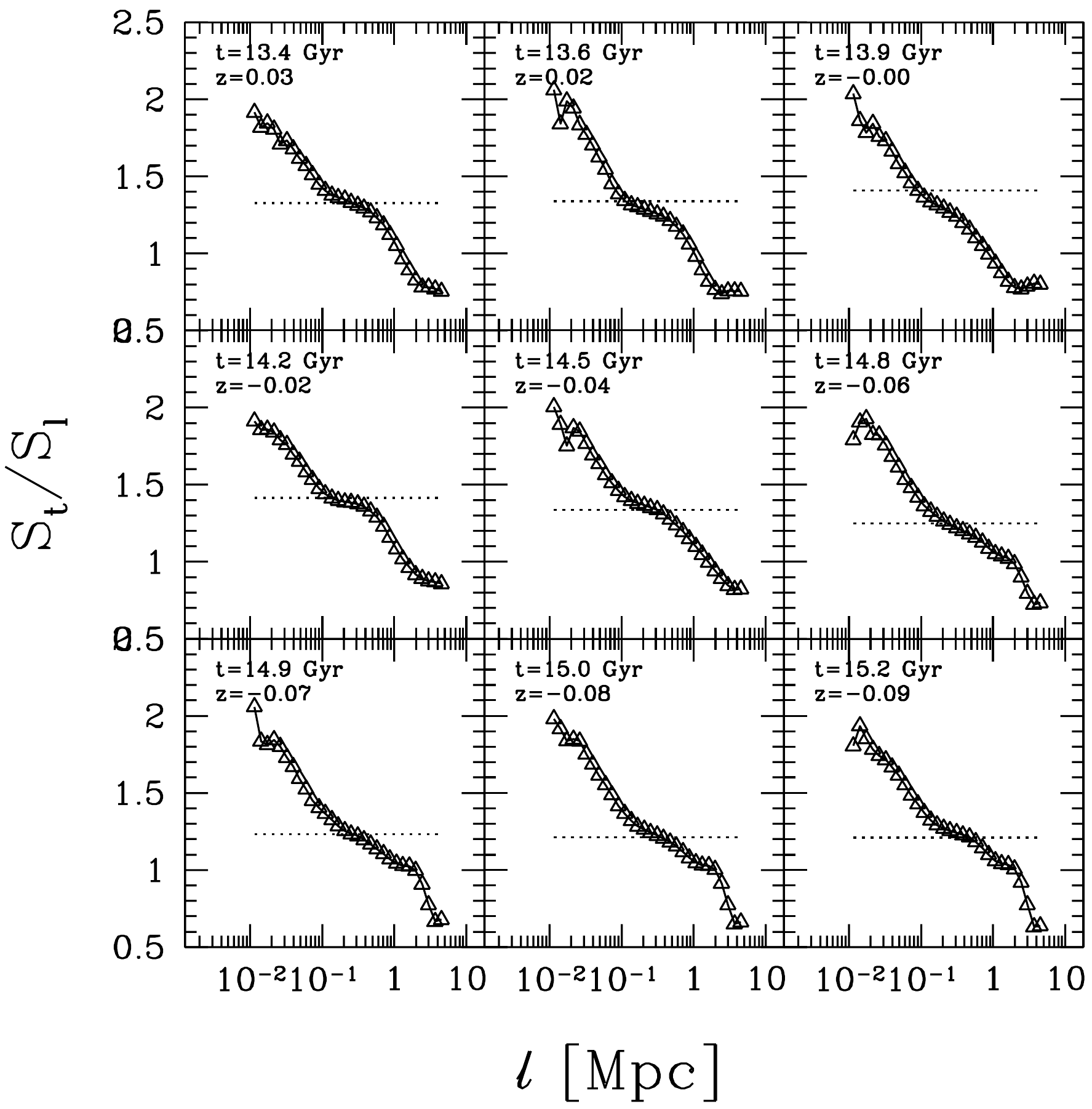}
\caption{
Ratio of transverse to longitudinal structure functions for the incompressible velocity component
 as measured from the simulation velocity data (triangle) versus expected value within the inertial range
 for fully-developed, steady-state, isotropic and homogeneous turbulence (dots).
\label{f5:fig}
}
\end{figure*}
\section{Results (I): Hydrodynamic Turbulence}\label{res:sec}
\subsection{Accretion History}\label{ah:sec}
The mass accretion history of this system is mildly irregular, due to the alternate dominance of 
substructure infall and smooth accretion, but remains quite uneventful until $z\approx 0.3$.
After this time the accretion rate increases significantly, culminating with two mergers events, a minor 
and a major one. 
The accretion history is reflected in the level of turbulent energy that is present in the ICM.
The turbulent velocity dispersion (see Eq.~\ref{ke:eq})
is reported as a function of time and redshift in Fig.~\ref{f1:fig}.
The solid line with open circles in the top panel correspond to the total velocity field, while the
dot-dashed and dashed lines correspond to the incompressible and compressional components,
respectively. Their ratio as a function of time is reported in the bottom panel of the same Figure.
At the earliest times in the Figure, $z\approx 1$, the turbulent velocity oscillates between several 
$10^2$ and a $10^3$ km s$^{-1}$.
It then settles at several hundred km s$^{-1}$ between $0.8 \lesssim z\lesssim 0.14$.
This is a apparently a time of relatively low mass accretion rate. Nevertheless, the drop in turbulent
energy is modest, indicating that turbulence is generated efficiently by mass accretion. 
This is not surprising with hindsight given that the dissipation of kinetic energy associated with 
accretion is very inefficient~\citep{2014ApJ...782...21M,2000ApJ...542..608M}. 
Finally, at $z\gtrsim 0.14$ the turbulent energy begins to ramp up. It peaks at 
$z\approx-0.4$ after reaching a secondary peak at $z\approx 0.05$.
The increase in turbulent energy is associated with two merger events, a minor one followed by a 
major one, described below in further details. In particular, the two peaks in the Figure
correspond to the time of core passage. Note that between $z\approx 0.3$ and  
$z\approx-0.4$ the turbulent energy has increased by a factor six.
The bottom panel of Figure~\ref{f1:fig} shows the fraction of turbulent energy in the compressional
component. This ratio is always smaller than one and oscillates between 10\% and 40\%.
There appears to be a correspondence between the higher values of this ratio and the higher
values of the turbulent energy (top panel). In particular, this ratio increases
during the two merger events.

The interval in which the mass accretion and turbulent energy are particularly high 
lies between $0.15 \lesssim z\lesssim -0.06$. 
Figure~\ref{f2:fig} shows the evolution of the merger during this specific period of time.
Each panel contains the density distribution in log scale on a plane passing approximately through the 
GC centre. Time, shown in the top-left corner of each panel, increases from top-left to bottom-right.
In the first panel (top-left) the main GC is visible at the centre, and is surrounded by 
a large structure to the left and two smaller structures, one to the top-right of (SubA),
and the other below (SubB), the main GC. 
The latter structure is actually not so visible because its centre lies on a slightly offset plane.
The first merger experienced by the GC is due to the accretion of the two substructures,
SubA and SubB. Since the substructure SubA is larger than SubB and its mass ratio to 
the GC is only 1/10, we are dealing with is a minor merger. 
It turns out, however, that SubA and SubB arrive at the GC core simultaneously and collide 
against each other (see top-centre and top-right panels). 
Despite the minor merger substantial turbulence
is stirred up in the ICM. The minor merger culminates at 12.9 Gyr (middle-left panel). 
Half a billion yr later the second merger takes place. In this case the mass ratio is 1/2.5, 
so this corresponds to a major merger event. The merger is in full swing by 13.9 Gyr 
(middle right panel) and core passage, i.e. crossover of the merging GC cores, occurs 
around 14.2 Gyr (bottom left panel). 
After that, the merging system expands inertially and the turbulence decays rapidly.

Using a standard approach~\citep{2005MNRAS.358..551B,2011MNRAS.412...49W,2010ApJ...718...60L}
for the major merger case we have also estimated the infall radial velocity of the 
substructure when located at the virial radius of the main GC. 
This velocity is 1642 km s$^{-1}$, to be compared to a virial velocity of 1024 km s$^{-1}$.
Thus this merger is quite energetic as 
such a high ratio of radial to virial velocity occurs only in less than 5\% of
mergers according to the statistical analysis of~\citet{2005MNRAS.358..551B}

It is interesting to note that the turbulent energy remains sustained for $\gtrsim 3$ Gyr, i.e. much 
longer than the duration of the major merger of $\simeq 1$ Gyr.
From the rapid decrease following its main peak (at $z\approx-0.4$) it is clear that the sustained
regime of turbulent energy is not due to sloshing following the first core passage.
Rather, the turbulent energy remains high for an extended period of time because the major merger
is not an isolated accretion event. Prior to that, a minor merger has occurred as already mentioned
and several other smaller substructures happened to accrete at that time. 
This fact is not occasional, but reflects the clustered character of matter distribution 
in the large scale structure of the universe. So rare events of major merger activity are likely 
accompanied by additional mergers of smaller structures, causing relatively high level of turbulence 
beyond the short duration of the major merger itself.
This point can only emerge when one studies the full time evolution of the turbulence in a
cosmological setting and can thus compare the level of turbulence during relaxed and 
unrelaxed conditions. In addition, it the details of mass accretion might be important when
making the connection between mergers and diffuse radio emission, because
a triple merger of even two minor substructure can rise significant of turbulence.
\begin{figure*}[t]
\centering
\includegraphics[width=0.5\textwidth]{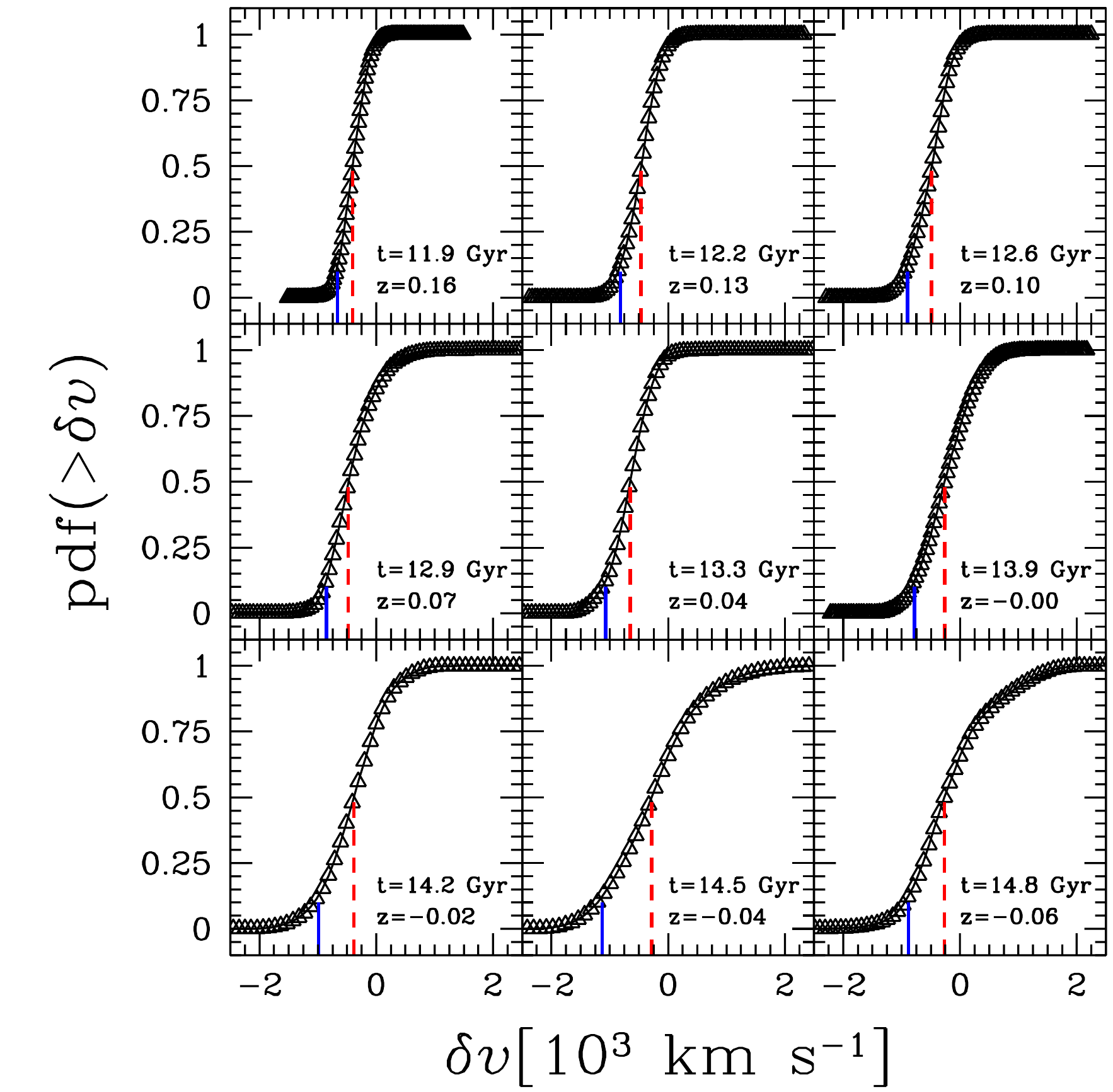}\includegraphics[width=0.5\textwidth]{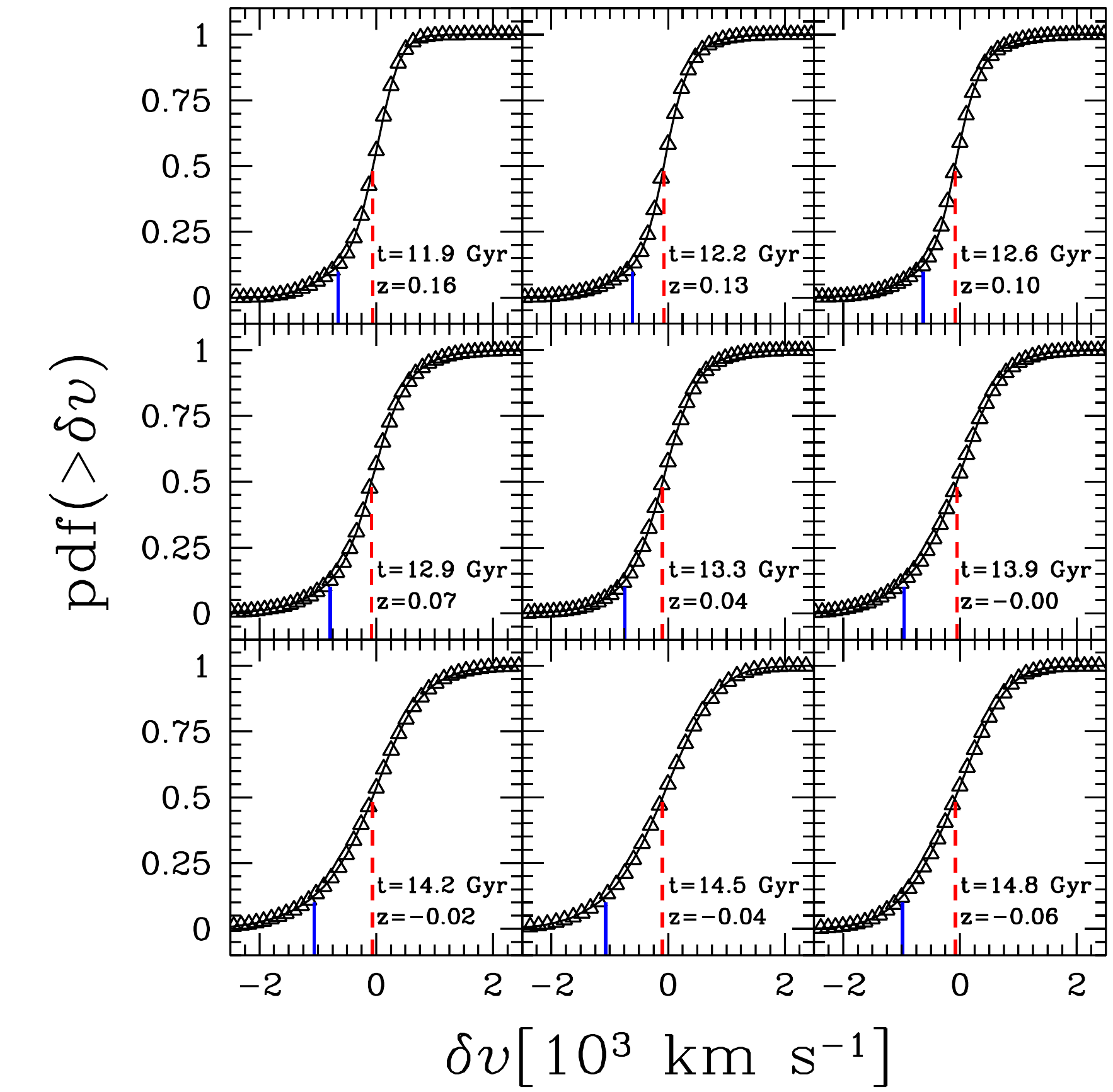}
\caption{
Cumulative probability density function for the compressional (left) and incompressible (right) velocity 
increment computed at separation given by the injection scale. Time in each panel is the same
as in the corresponding panels of Figure~\ref{f2:fig}. The dashed (red) vertical line marks the median 
value while the solid (blue) vertical line the 12.5th percentile.
\label{f6:fig} }
\end{figure*}
\subsection{Time Dependent Statistics}\label{tds:sec}
In this Section we discuss the evolution of the statistical properties
of the turbulence focusing on the time interval $0.28 \lesssim
z\lesssim -0.09$, which encompasses both the extended period of high
turbulent energy and the time prior to it when the turbulence level
was at a minimum.

We begin with second order structure functions shown at different
times in Figure~\ref{f3:fig}, for both the incompressible (triangles)
and compressional (dots) velocity component.  The structure functions
are computed within a region of one Mpc from the GC centre
(see~\ref{analysis:sec}).  The numerical aspects concerning the
modelling of turbulence in structure formation simulations are fully
discussed in~\citet[][and references therein]{2014ApJ...782...21M}.
In general the high resolution allows us to resolve at least one
decade of scales in the inertial range of the turbulent cascade: from
the injection scale, typically a Mpc, to the scales affected by
numerical viscosity, roughy 32 resolution element, i.e. 0.1 comoving
Mpc.

We have thus identify the inertial range of the structure functions
and computed the corresponding slope and normalisation.  These are
represented by the solid lines superposed to the triangles and dots.
Based on the flattening of the structure functions, we have then
identified for each velocity components the injection scale, $L$ (red
{\it vertical} solid lines in Figure~\ref{f3:fig}).  By extrapolating
the velocity from the inertial range to $L$, we finally obtain the
asymptotic velocity increment (red {\it horizontal} solid lines in
Figure~\ref{f3:fig}). 
Figure~\ref{f3:fig} shows that it is fairly straightforward to
identify such quantities. In particular the inertial range is rather
well defined. 
Occasionally, however, the inertial range of
the structure functions of the compressional component 
appears as a broken power law, indicating
departure from steady state conditions.
In this case, our extrapolation
procedure leads to velocity estimates slightly offset from the dotted
curve itself. This is unavoidable consequence of our simplified
representation of the structure functions with few parameters, in the
case of departure from simple power-law. The theory/models to which
this results will be applied in Sec. 4 do not warrant the use of
additional parameters on the basis of accuracy, therefore we will
tolerate the inaccuracy. In fact, these issues arises primarily during
pre and post-merger phases, not during the crucial stages of the
merger itself (around core passage), so it does not affect the results
discussed in Section~\ref{resII:sec}. 
In any case our estimates remain conservative and, if anything,
overestimate the turbulent kinetic energy.
Finally note that, while it would be
possible to evaluate the outer scale and velocity independently of the
inertial range, we think this would be even less representative of the 
turbulent cascade.

According to Figure~\ref{f3:fig} the structure functions
change both in normalisation and slope as a function of time.  This
happens in response to the changing mass accretion rate and merger
events, which obviously changes the level of turbulent energy in the
ICM.  Figure~\ref{f4:fig} shows in detail the time dependence of the
structure functions' slopes in the inertial range. The structure
function is always steeper for the compressional than for the
incompressible velocity. In addition its changes are considerably
larger and on shorter timescales. After t=10.6 Gyr the turbulence
starts to settle to a low level (see Sec.~\ref{ah:sec}).  Both
structure functions steepen. The slopes tend to 0.8 and 1.5 for the
incompressible and compressible components, respectively (see
Figure~\ref{f4:fig}). After t=12 Gyr the minor merger discussed above
begins to take place and both structure functions become progressively flatter and
with a larger normalisation. Their slopes become comparable at the
peak of the merger, although the compressional component remains
always steeper.  During the short time separating the first (minor)
and second (major) merger, the compressional turbulence decays
quickly, its structure function steepens and breaks, in contrast to
the slow evolution of the incompressible counterpart. The decay of the
turbulence is interrupted by the occurrence of the second merger. For
both components, the normalisation increases again while the slope
flattens. The two structure functions at t=14.2 Gyr are almost
parallel. After core passage the compressional turbulence again decays
rather quickly, helped by the re-expansion of the ICM driven by the
departing motions of the merging structures. In contrast, the
incompressible component decays very slowly while its spectrum appears
to get flatter, as if receiving additional power at intermediate
scales.

In Figure~\ref{f5:fig} we plot as a function of spatial separation and for the same time sequence
as Figure~\ref{f3:fig}, the ratio of the transverse 
to longitudinal second order structure functions of the incompressible velocity component.
For fully developed, incompressible, isotropic and homogeneous turbulence, 
in the inertial range of the structure functions this ratio is expected to be~
\citep{deKarman:1938dr,LandauLifshitz6}, 
\begin{equation}\label{vKH:eq}
\frac{S_{s,t}^{(2)}}{S_{s,l}^{(2)}} = 1 +\half\zeta_2.
\end{equation}
The predicted value is also plotted as a dotted line in each panel of Figure~\ref{f5:fig}, 
according to the time dependent slope of the structure functions. Note that the dissipative
effect of  numerical viscosity appears below 0.1 Mpc and is strongly 
anisotropic, i.e. it preferentially damps the longitudinal velocity increments~\citep{2014ApJ...782...21M}. 
In any case, the conditions for fully developed turbulence appear to be approximately fulfilled
above the numerical dissipation scale and below the injection scale $\sim $ Mpc.
This is sufficient to warrant our approach based on the evaluation of the total structure functions
in Eq.~\ref{tsf:eq}. While the quality of such fulfilment is obviously not perfect, this is not a numerical
artefact but due to the nature of the ICM. Interestingly, departures from steady-state conditions
appear to be worse at times when the turbulence is weak, $z\gtrsim 0.2$ and $z\lesssim -0.06$,
and immediately before core passage, $t=12.9$ and $t=13.9$.

Finally, in Figure~\ref{f6:fig} we present the cumulative distribution function of the velocity increment 
for a separation scale corresponding to the injection scale (where the velocity increments are 
largest). The left and right panels are for the 
compressional and solenoidal velocity components, respectively. The dashed (red) vertical line marks 
the median value of the velocity whereas the solid (blue) vertical line the 12.5th percentile. Time 
in the left and right panels is the same as in the corresponding panels of Figure~\ref{f2:fig}, i.e. is
around the period of enhanced mass accretion.
As for the compressional component we find that it is characterised by a negative median velocity 
increment, i.e. $\langle\Delta\vvec\rangle_M<0$. This implies a systematic 
convergence of the compressional motions. Such motions can produce heating from adiabatic 
compression (AdC) of the particles and is important in view of the acceleration mechanisms 
discussed in the next Section. The heating rate can be approximated as
\begin{equation}\label{gad:eq}
\Gamma_p=\frac{1}{p}\frac{dp}{dt}=-\frac{1}{3}\nabla\cdot\vvec\approx -\frac{\langle\Delta\vvec\rangle_M}{3L}
\end{equation}
We anticipate that while the contribution from this mechanism is comparable to others already 
proposed in the literature, it is relatively slow and does not change the qualitative picture.
Note that no such motions appear in the incompressible velocity component, as expected.
Note also how the distribution of the velocity changes with time and in particular broadens
in connection with mergers events. However, there is no sign of significant regions of strongly 
enhanced motions. In fact, the 12.5 percentile value remains remains typically 
$\lesssim 10^3$ km s$^{-1}$, comparable to the velocity at the outer scales in Figure~\ref{f3:fig}.
This is also of interest from the viewpoint of particle acceleration in the ICM discussed in the next 
Section.

\section{Results (II): Stochastic Particle Acceleration}
\label{resII:sec}
In the following we use the above results concerning the statistical properties of the turbulence
to compute the acceleration rates predicted by models based on the interaction of particles with
turbulence generated waves. 
In particular we consider the acceleration processes referred to as 
non-resonant mechanism~\cite[NR;][BL7]{1988SvAL...14..255P} and 
transient-time-damping~(TTD; BL7, BL11).
These models have received particular attention in connection with the presence of diffuse 
radio emission in galaxy clusters.
In brief, TTD is a second order Fermi mechanism in which particles in phase-resonance 
with the fast magneto-sonic waves are reflected through the magnetic 
mirror effect~\citep{1976JGR....81.4633F,1991ApJ...376..342M}.
In the non-resonant mechanism~\citep{1988SvAL...14..255P}, suprathermal particles are 
accelerated by stochastic fluctuations in the velocity divergence according to the adiabatic process 
$\frac{dp}{dt}=-\frac{p}{3}\nabla {\bf u}$. 
Here the correlation scale of the compressional turbulence, $\ell$, is much larger than the particle
scattering mean-free-path, i.e. $\ell\gg\ell_{mfp}\simeq 3D/c$, where $D$ is the diffusion coefficient, 
assumed homogeneous and isotropic. This process is of the same nature as, and superposes 
independently to, the adiabatic heating described in Eq.~(\ref{gad:eq}).
In both the above stochastic models the particles interact with 
compressional fast magneto-sonic waves, which in the high-$\beta$ ICM plasma effectively correspond
to acoustic waves. The acceleration rates are sensible to 
a number of assumptions concerning the microphysics of the ICM plasma. 
In particular the dynamics of the cascade of fast magneto-sonic waves and the resulting cutoff scale.
In order to expose how such dependencies emerge in our results, in the following 
we first briefly review the formulae expressing the acceleration rates that we use below.

The evolution of the distribution function of relativistic particles, $f$, 
is described by the transport equation 
which in the case of isotropic diffusion reads~\cite[e.g.][]{1987PhR...154....1B}
\begin{equation}\label{lineq:eq}
\frac{\partial f}{\partial t}+{\bf v}\cdot\nabla f-D_{xx}\nabla f-\frac{1}{p^2}\frac{\partial}{\partial p}\left[p^2\left(b(p)f+D_{pp}\frac{\partial f}{\partial p}\right)\right]=0.
\end{equation}
Here $D_{xx}$ and $D_{pp}$ are the Fokker-Planck diffusion coefficients in configuration
and momentum space, respectively and, $b(p)\equiv-dp/dt$, is the particle's total energy loss 
rate~\citep[see e.g.,][]{2001ApJ...562..233M}. The particle propagation in momentum space can be described
as a combination of diffusion governed by the diffusion coefficient, $D_{pp}$, 
and advection at the rate
\begin{equation}\label{gamma_p:eq}
\Gamma_p\equiv \frac{\dot p}{p}=p^{-3}\frac{\partial}{\partial p} (p^2 D_{pp}) = 4 \frac{D_{pp}}{p^2}.
\end{equation}
The last equality is based on the fact that for the processes considered below $D_{pp}\propto p^2$.
Thus advection in momentum space is linear. This is also the case for adiabatic 
compression (Eq.~\ref{gad:eq}), which however produces no diffusion. Finally, note that $b(p)$ 
also acts as a advection term, albeit a nonlinear one. In the following we begin reviewing the 
expressions for the diffusion coefficients according to the processes of interest.

\subsection{Diffusion Coefficients} \label{dpp:sec}
The diffusion coefficient for the TTD process~(BL7 and references therein)
can be written, after some manipulation, as
\begin{gather} \label{dpp_ttd:eq}
D_{pp}(p)= p^2\frac{\pi I_\theta\left(\frac{c_s}{c}\right)}{8c} 
\langle k\rangle_W\, \langle(\delta\vvec_c)^2\rangle,
\end{gather}
where
\begin{gather}  \label{kE:eq} 
\langle k\rangle_W\equiv\frac{1}{\langle(\delta\vvec_c)^2\rangle}\int^{k_c} dk k W(k) \simeq \frac{\zeta_2}{1-\zeta_2} k_L \left(\frac{k_c}{k_L}\right)^{1-\zeta_2}.
\end{gather}
In addition, the total energy is $W(k)\approx 2 E(k) $ and is normalised to the velocity dispersion (twice the kinetic energy,~see Section~\ref{analysis:sec}), 
$k_L\equiv2\pi/L$, and $k_c$ is the cutoff mode of the spectrum of fast magneto-sonic waves.
The integral $I_\theta(x)\equiv\int_0^{\arccos(x)} d\theta\frac{\sin^3\theta}{|\cos\theta|}\left[1-\left(\frac{x}{\cos\theta}\right)^2\right]^2\approx 5$ for a typical sound speed of 10$^8$ cm s$^{-1}$, i.e. $x=c_s/c\simeq 1/300$.

For the non-resonant mechanism~\cite{1988SvAL...14..255P}, again after some manipulation,
we write
\begin{gather}\label{dpp_nr:eq}
D_{pp}=p^2\frac{2}{9}\zeta_2\frac{\langle (\delta\vvec_c)^2 \rangle}{D}
\left(\frac{k_LD}{c_s}\right)^{\zeta_2}
\int_{k_LD/c_s}^{k_cD/c_s} d\xi
\frac{\xi^{1-\zeta_2}}{1+\xi^2},
\end{gather}
where we assume $D=\frac{2\pi}{3}c/k_c$, i.e. we relate the mean free path to the cascade cutoff (BL7).
This is a strong simplification and refined assumptions shall be explored in the future.
However, as we shall see, as long as $k_c\gg k_L$, this case remains
weakly  dependent on such fine details.
In any case, in general in both cases, 
the diffusion coefficient depends on the details of the cascade of fast magneto-sonic 
waves, through the parameters $\zeta_2$ and $k_{c}$.
Thus in the next Section we summarise how these quantities are determined. 
Arguably, the other quantities entering the diffusion coefficients, such 
as the sound speed, $c_s$ (which also determines $I_\theta$), the injection 
scale $L$ and the corresponding velocity dispersion $\langle(\delta\vvec_c)^2\rangle$ 
can be obtained with reliable approximation from the hydrodynamic simulation. 

\subsection{Cascade Models}\label{camo:sec}

In~BL7 the cascade of fast magneto-sonic waves is based on Kraichnan's picture.
The cascade rate is determined by the Alfv\'en speed, $v_A=B/\sqrt{4\pi\rho}$ 
and is $\tau_A^{-1}=v_Ak$, rather than the eddy turnover time, 
$\tau_{NL}^{-1}=k^{3/2}E(k)^{1/2}$~\citep{1990JGR....9514881Z}. Energy transfer 
to higher modes is therefore faster, resulting in a flatter spectrum, than Kolmogorov's.
In particular, given the injection spectrum of turbulent energy per unit mass 
\begin{equation}
I(k)=I_0\delta(k-k_{L}),\quad I_0=C_Wk_L\frac{\langle(\delta\vvec_c)^2\rangle ^2}{c_s},
\end{equation}
the equilibrium spectrum for the total energy is
\begin{equation}
W(k) \approx 2E(k)=
\left(\frac{2C_W}{7}\right)^\half \frac{\langle(\delta\vvec_c)^2\rangle}{k_L}\left(\frac{k}{k_L}\right)^{-\frac{3}{2}}.
\end{equation}
By normalising $k_LW(k_L)$ to the velocity dispersion we infer $C_W=7/2$
(comparable to 81/14 in BL7; note, however, that by normalising instead the 
integral of the total energy one gets $C_W=7/8$). 
Kraichnan's scenario, however, 
should apply when $\tau_A^{-1}\gg\tau_{NL}^{-1}$~\citep{1990JGR....9514881Z}, 
i.e. below the Alfv\'en scale at $k>k_A=k_L\langle(\delta\vvec_c)^2\rangle^{n/2}/v_A^n$, where
$n=3,$ respectively 2, for Kolmogorov's and Burgers cases. So it is not clear that this it is the 
dominant mechanism for the cascade of compressional, fast magneto-sonic waves in the high-$\beta$ ICM. 
Magneto-sonic waves could partly dissipate through weak shocks. 
This is indeed what happens in our hydrodynamical model. In this case the 
equilibrium spectrum is considerably steeper, similar to the case of Burgers' turbulence~\citep{2014ApJ...782...21M}. We will consider this further in our estimates of the acceleration
rates below.

Another important quantity entering the estimates of the acceleration rates is the cutoff 
scale of the turbulent cascade. At this scale the cascade rate equals the dissipation rate, 
so it depends on both the dissipation mechanism and the cascade model.
In BL7 the main damping mechanism of the cascade is TTD of thermal electrons. 
This assumes that the thermal plasma is collision-less so the long mean-free-path 
of the thermal particles warrants the resonant interaction with the fast magneto-sonic waves.
The azimuthally averaged dissipation rate is
\begin{gather}
\frac{\langle\Gamma_{\rm TTD}\rangle}{k}
\approx \left(\frac{3\pi m_e}{20 m_p}\right)^\half c_s 
\left\langle\frac{\sin^2\theta}{\cos\theta}e^{-\frac{5m_e}{3m_p\cos^{2}\theta}} \right\rangle_\theta,
\end{gather}
where $\langle\cdots\rangle_\theta\sim 2.7$.
By equating the Kraichnan's cascade rate and the above dissipation rate leads 
to~(BL7)
\begin{gather}\label{kckttd:eq}
k_c=\frac{81}{14}\left(\frac{I_0}{c_s}\right) \left[\frac{\langle\Gamma(k)\rangle}{k}\right]^{-2} 
\approx A k_L\frac{\langle(\delta\vvec_c)^2\rangle^2}{c_s^4},
\end{gather}
with $A=(\frac{270C_Wm_p}{7\pi m_e})\left\langle\frac{\sin^2\theta}{\cos\theta}e^{-\frac{5m_e}{3m_p\cos^{2}\theta}} \right\rangle_\theta^{-2}\approx 11000$, with the above choice of $C_W$.

If shock dissipation is effective and the cascade is steeper as in Burgers' model,
the effective energy transfer rate is slower and the cutoff scale correspondingly larger.
One way to roughly estimate the decay rate is to write 
$\epsilon_\ell=\langle(\delta\vvec_c)^2_\ell\rangle/\tau_\ell=\langle(\delta\vvec_c)^3_\ell\rangle/\ell$ so that 
\begin{gather}
\tau_\ell^{-1} \approx \frac{\langle(\delta\vvec_c)^3\rangle}{\langle(\delta\vvec_c)^2\rangle}L^{-1}\left(\frac{\ell}{L}\right)^{\zeta_3-\zeta_2-1}.
\end{gather}
A similar expression can also be found using more rigorous arguments~\citep{LandauLifshitz6}.
Equating this rate to the TTD rate gives
\begin{gather}\label{kcbttd:eq}
k_c\approx 
\left( 2\pi\frac{\langle\Gamma_{\rm TTD}\rangle}{k}\frac{\langle(\delta\vvec_c)^2\rangle}{\langle(\delta\vvec_c)^3\rangle}\right)^{-\frac{1}{\zeta_3-\zeta_2}} k_L
\end{gather}
which is typically much larger than the value of Eq.~(\ref{kckttd:eq}).
Finally, a number of authors have suggested that the effective mean-free-path of the thermal particles
could be substantially reduced due to the action of small scale 
instabilities~\citep{Parker58,Ginzburg79,Schekochihin05,Schekochihin08}.
This prevents the resonance of the thermal electrons with the fast magneto-sonic waves 
suppressing the associated TTD mechanism. This case has been addressed in BL11. 
The main consequence is that the relativistic particles become the main mechanism for 
the dissipation of fast magneto-sonic waves through the TTD process.
As a result, dissipation of fast magneto-sonic waves becomes less efficient,
their cascade extends to much smaller scales and acceleration of relativistic
particles becomes much more efficient. Following BL7, 
we find the azimuthally averaged dissipation rate due to relativistic particles
\begin{equation}\label{gacrttd:eq}
\frac{\left\langle\Gamma_{CR}\right\rangle}{k}=\frac{\pi I_\theta}{8}\frac{q\epsilon_{CR}}{\rho c}
\end{equation}
where $q\simeq 4.5$ is the log-slope of the cosmic-ray distribution function,
and $\epsilon_{CR}$ is the cosmic-ray energy density. Then the cutoff mode 
is given by Eq.~(\ref{kckttd:eq}) but with the above dissipation rate, namely
\begin{equation}\label{kccttd:eq}
k_c\approx \frac{81}{14}C_W\left(\frac{8}{\pi I_\theta qx_{CR}}\right)^2k_L\frac{\langle(\delta\vvec_c)^2\rangle^2}{c_s^6}c^2,
\end{equation}
where, $x_{CR}=\epsilon_{CR}/\rho c_s^2$ is the ratio of cosmic-ray to thermal energy.
Then the diffusion coefficient for the NR process is given by Eq.~(\ref{dpp_nr:eq}) with the
above value of $k_c$ while the diffusion coefficient associated to the TTD mechanism 
(Eq.~\ref{dpp_ttd:eq}) becomes
\begin{equation}\label{dpp_c:eq}
D_{pp}=p^2C_D\frac{I_0\xi}{\epsilon_{CR}}=p^2\frac{C_DC_W\xi}{x_{CR}}k_L\frac{\langle(\delta\vvec_c)^2\rangle^2}{c_s^3},
\end{equation}
where $C_D=(2/q)(81/49)^\half\approx0.57$ with $q\simeq 4.5$.
In the above expression we have introduced an additional factor $\xi$ to account for the 
effectiveness of the above microscales instabilities to render the plasma highly collisional.
For example, high temporal or spacial intermittency of the instabilities could still enable some 
damping of the compressible waves via TTD by thermal particles.

In any case, this scenario boosts significantly the efficiency of the TTD mechanism, while only marginally that
of NR. It is worth mentioning that in BL11, Kraichnan's model is assumed for the cascade. As already
noted, however, the cascade could partly dissipate also through weak shocks, reducing 
dramatically the energy in the cascade available for, and the efficiency of, particle acceleration 
regardless of the plasma collisionality. In this case the 
the diffusion coefficient would be given again by Eq.~(\ref{dpp_ttd:eq}) with cutoff scale as in
Eq.~(\ref{kcbttd:eq}) but with the dissipation rate due to cosmic-rays given in 
Eq.~(\ref{gacrttd:eq}). This results in a larger cutoff mode than the corresponding collisionless case.
However, because $\zeta_2\approx 1$, this larger value has little effect on $\langle k\rangle_W$
(see Eq.~\ref{dpp:sec}), and hence on $D_{pp}$, just as in the BL7b case. 
Therefore we will dismiss this specific case in the following discussion, keeping in mind that it gives 
similar results to the BL7b case.

\begin{figure*}[tp]
\centering
\includegraphics[width=0.5\textwidth]{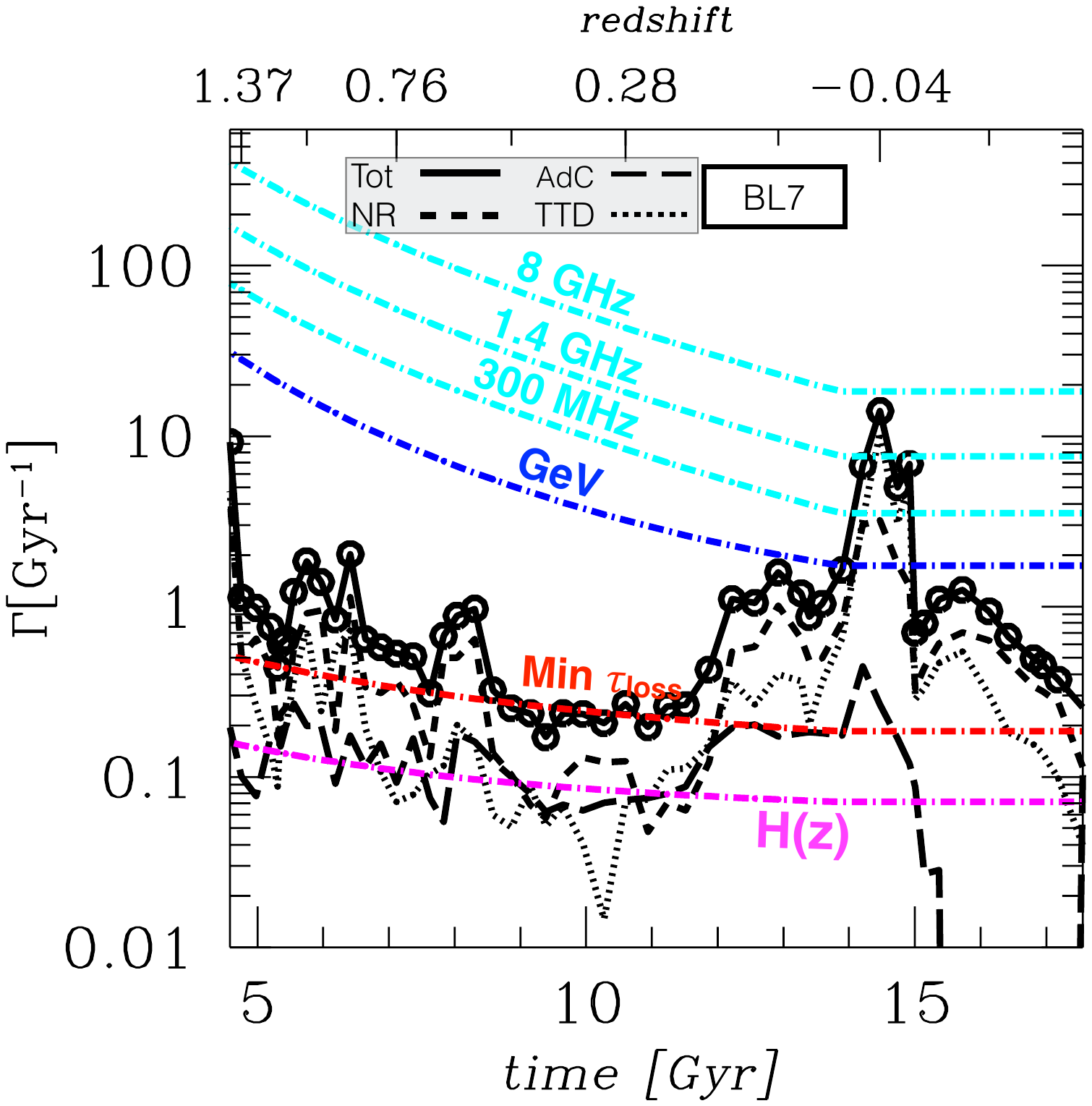}\includegraphics[width=0.5\textwidth]{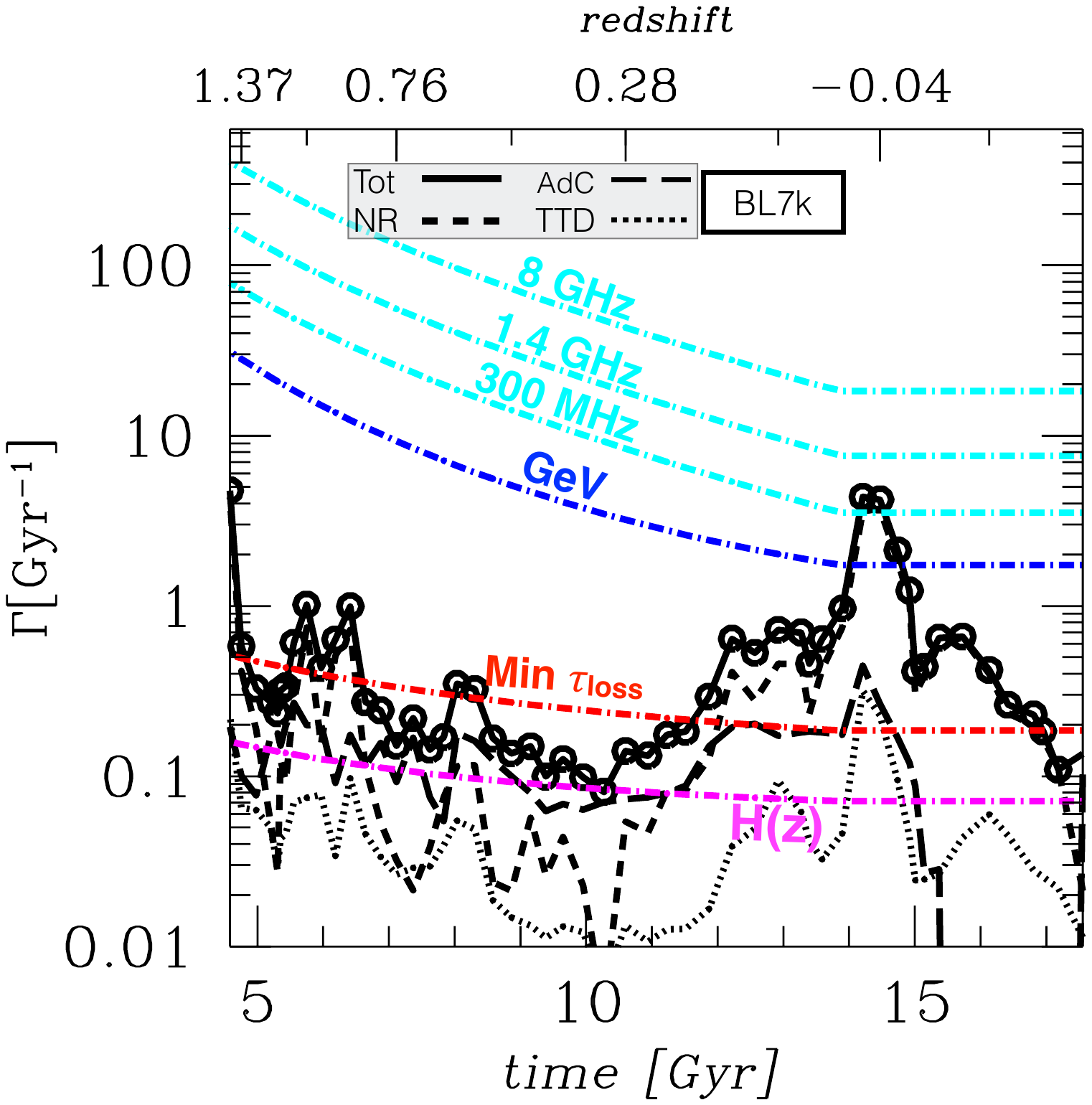}
\includegraphics[width=0.5\textwidth]{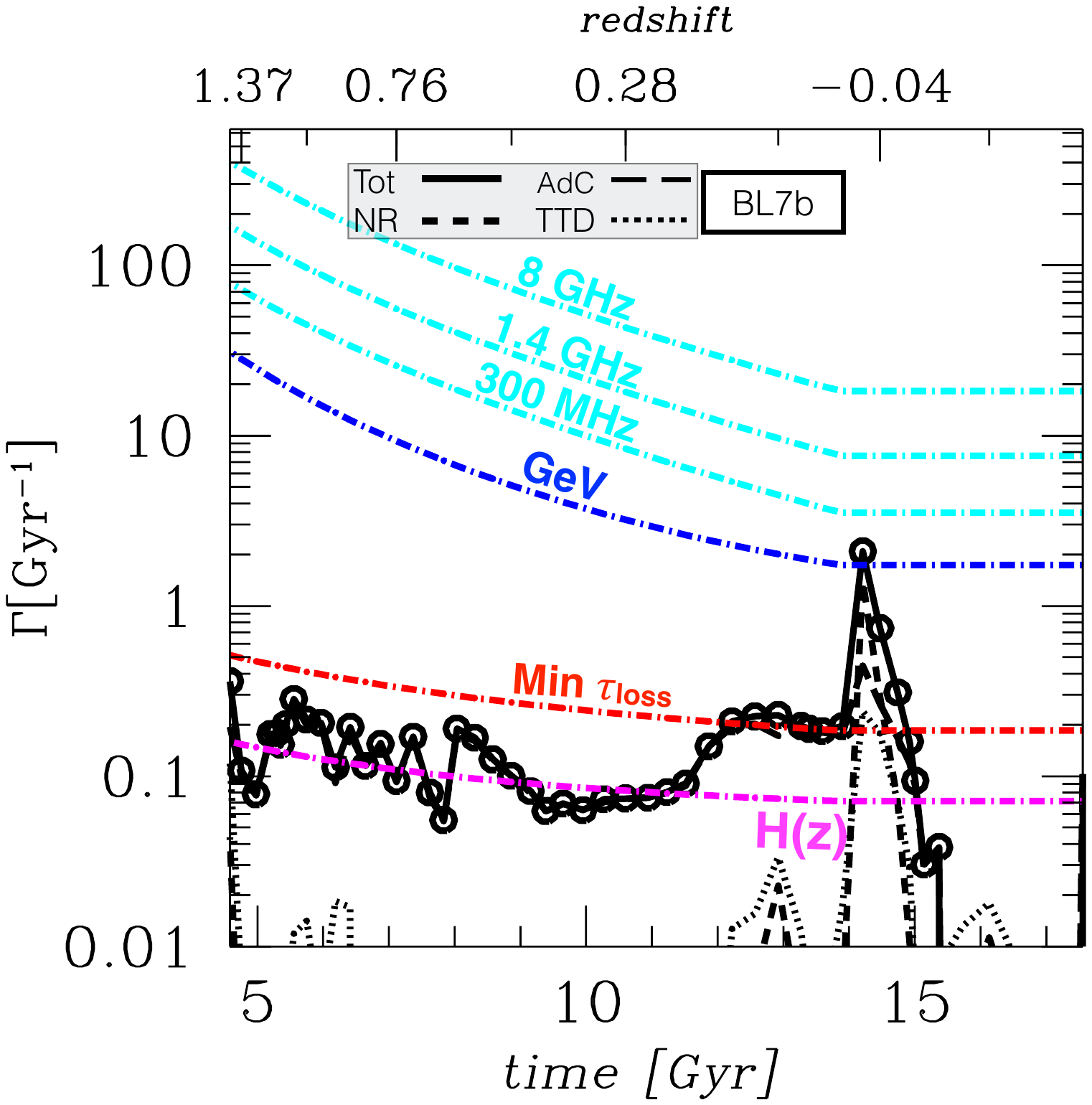}\includegraphics[width=0.5\textwidth]{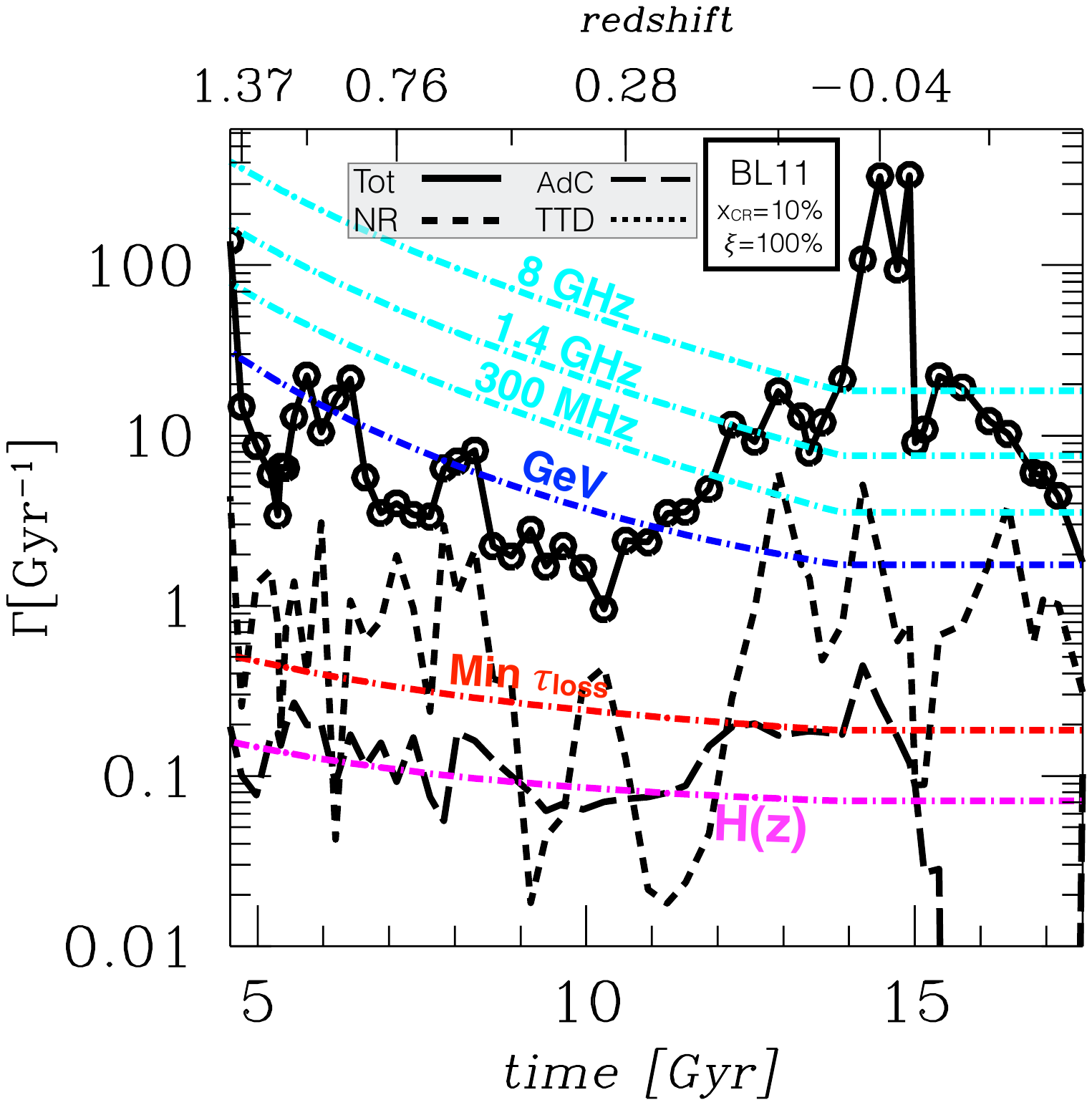}
\caption{
Acceleration rates due to the NR (short-dash), TTD (dotted) mechanisms and AdC (long dash) 
as a function of time in units of Gyr (bottom axis) or redshift (top axis).
The four panels correspond to the different physical scenarios for the microphysics of BL7 (top left),
BL7k (top right), BL7b (bottom left), BL11 with $x_{CR}=10\%$ and $\xi=100\%$ (bottom right) and their 
total sum (solid line with open dot symbols).
The cyan curves correspond to the minimum energy loss rate of an electron 
emitting synchrotron radiation at 300 MHz (lower), 1.4 GHz (middle) and 8 GHz (upper).
Also shown are the minimum energy losses suffered by a suprathermal particle in the ICM for a 
number density of free electrons $10^{-3}$ cm$^{-3}$ and magnetic field strength of 10$^{-6}$ G 
(red curve); the energy loss rate for a GeV electron (blue curve); and the Hubble rate (magenta).
\label{f7:fig}
}
\end{figure*}
\subsection{Momentum Advection Rates}\label{adr:sec}

\begin{figure}[t]
\centering
\includegraphics[width=0.45\textwidth]{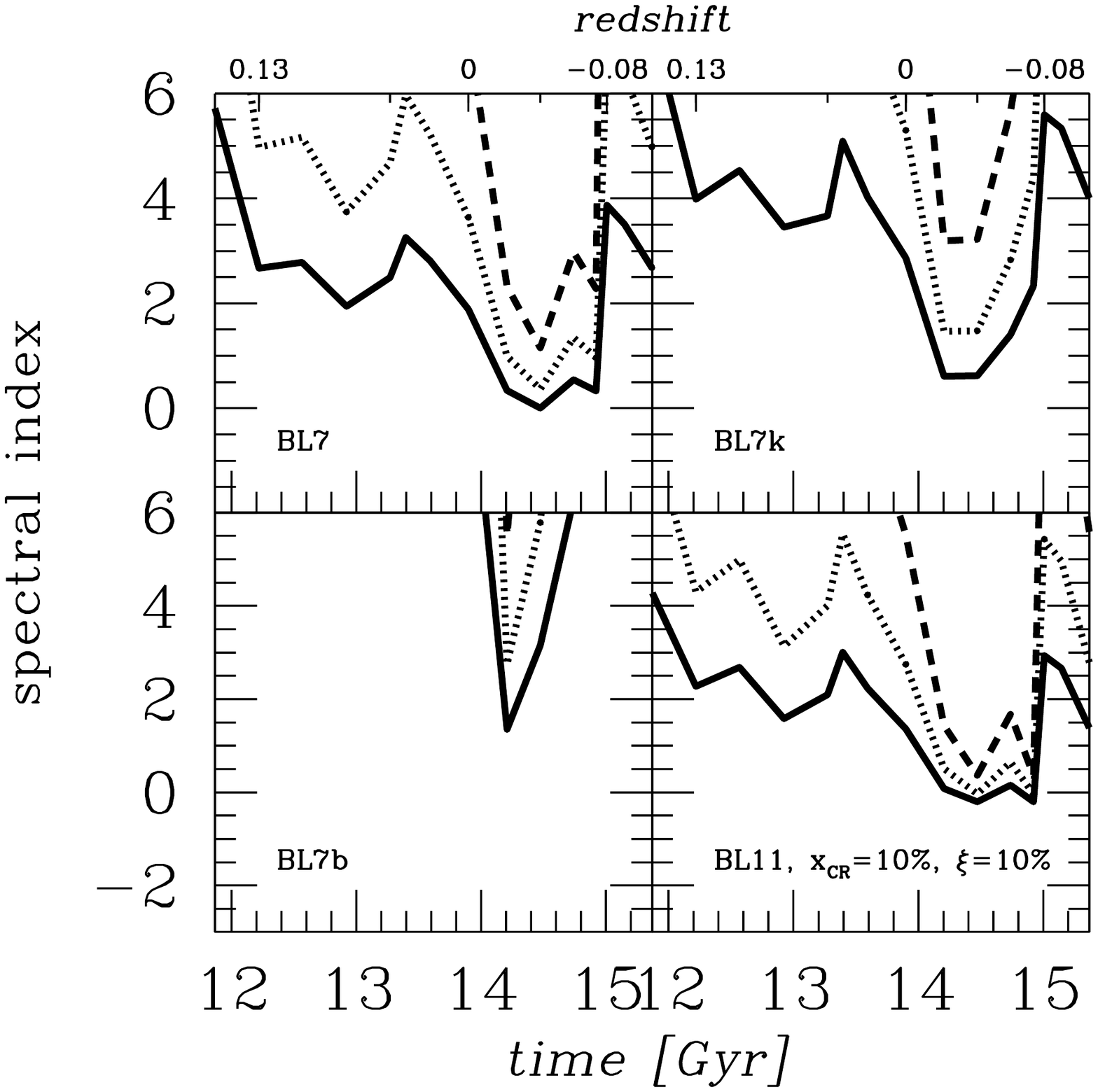}
\caption{Spectral index 
for emission at 300 MHz (solid line), 1.4 GHz (dotted line)
and 8 GHz (dashed line) as a function of time in Gyr 
(bottom axis) and redshift (top axis). The top left panel is for BL7, the top right  for BL7k
and the bottom left for BL7b (same as the corresponding top panels in Figure~\ref{f7:fig}).
The bottom right panel is for BL11 with $x_{CR}=10\%$
and an additional intermittency factor, $\xi=10\%$.
\label{f8:fig}
}
\end{figure}

Using the statistical properties of the compressional turbulence (Figure~\ref{f3:fig} and ~\ref{f6:fig}) 
we now compute the advection rates in momentum space (Eq.~\ref{gamma_p:eq}) for the processes
discussed in the previous Sections. Remember that, since the diffusion coefficients in these cases turn
out proportional to $p^2$, the advection timescales are momentum independent.
The advection rates are plotted in Figure~\ref{f7:fig} in units of inverse Gyr, 
as a function of time in units of Gyr (bottom x-axis), and redshift (top x-axis) for
TTD (dotted line), NR (short-dashed line), AdC (long-dashed line), together with
their total sum, Tot (solid line and open symbol). The four panels correspond to four different 
scenarios for the microphysics discussed in the previous Section.
In particular, the top-left panel corresponds to the BL7 case, meaning that the cascade of
compressional modes is based on Kraichnan's model with cutoff given by Eq.~(\ref{kckttd:eq}).
The case of the top-right panel (BL7k) is the same as BL7 except for the cascade's slope, which is 
taken from the hydrodynamic model. Of course this is not self-consistent but provides 
a useful intermediate step to the results of the bottom left panel (BL7b) where the full cascade is 
based on the hydrodynamical model with cutoff given by Eq.~(\ref{kcbttd:eq}). Finally, 
the bottom right panel assumes the scenario in BL11,
namely Kraichnan's cascade model with cutoff given by Eq.~(\ref{kccttd:eq})
motivate by an effective high collisionality of the ICM plasma.
The acceleration rates are compared to the minimum energy loss rate, $\Gamma_{\rm min}$, 
suffered by a relativistic electron emitting synchrotron radiation at the following frequencies: 300 MHz 
(lower dot-dashed cyan curve), 1.4 GHz (middle) and 8 GHz (upper). These are the typical 
frequencies at which diffuse radio emission from the ICM is observed.
At a given frequency, $\nu$, and redshift, the minimum energy loss rate is
\begin{equation}\label{Gamin:eq}
\Gamma_{min}(\nu)=a\,(1+z)^\frac{7}{2}\left(\frac{\nu}{\nu_{\rm CMB}}\right)^\half\quad {\rm Gyr}^{-1},
\end{equation}
occurring in a magnetic field strength
\begin{equation}\label{Bmin:eq}
B_{min}=\frac{B_{\rm CMB}(z)}{\sqrt{3}}=\frac{B_{\rm CMB}(0)}{\sqrt{3}}(1+z)^2,
\end{equation}
where $a=6.2\times 10^{-4}$, $\nu_{\rm CMB}=eB_{\rm CMB}/(2\pi m_ec)\approx 9$Hz and 
$B_{\rm CMB}$ is the CMB energy density expressed in terms of magnetic field strength.
Also, the momentum of the radiating particles is
\begin{equation}\label{pmin:eq}
p_{min}=(1+z)^{-\half}\left(\frac{\nu}{\nu_{\rm CMB}}\right)^\half m_ec.
\end{equation}
For reference in Figure~\ref{f7:fig} we also show the minimum energy losses suffered by a 
suprathermal particle in an ICM with number density of free electrons $n_e=10^{-3}$ cm$^{-3}$ 
and magnetic field strength $B=10^{-6}$ G (red curve), the energy loss rate for a GeV electron 
(blue curve), and the Hubble rate (magenta).

In the BL7 scenario (top-left panel) the total advection rate is basically always 
above the minimum loss rate curve, meaning that is it sufficient to maintain 
the bulk of a low energy (ca 100 million eV) but suprathermal population of electrons.
In this scenario the AdC process contributes comparably to TTD and NR up to 12 Gyr,
is almost negligible during the merger event and actually contributes to the cooling of the
particles after core passage.  Note that the NR advection rate increases
significantly during the 3 Gyr long period of high mass accretion rate described in Section~\ref{ah:sec}
while the TTD only during the major merger itself. 
In addition, the two peaks in the total acceleration rate correspond to core passage of the 
minor and major merger events. During the major merger in particular, in this scenario 
the total advection rate reaches values close to the minimum energy losses of a 1.4 GHz emitting electron.
When the slope of the cascade is modified from Kraichnan to 
the value of the hydrodynamic model (top-left) the TTD rate drops dramatically,
while the NR rate is almost unaffected. The drop of TTD rate is ascribed to the sensitive dependence
of the quantity $\langle k\rangle_W$ on $\zeta_2$ in Eq.~(\ref{kE:eq}). In particular, $\langle k\rangle_W$
and, as a consequence, $D_{pp}$, become progressively smaller for  $\zeta_2$ progressively larger 
than $\half$,
which is what happens when Kraichnan's slope is replaced by the one from the hydrodynamic model.
Note that this effect occurs regardless of the plasma collisionality, i.e. TTD is generally suppressed when 
$\zeta_2\rightarrow 1$.
On the other hand, the NR rate is less sensitive to changes in $\zeta_2$ and, as long as $k_c\gg k_L$,
it becomes almost independent of $k_c$ as $\zeta_2\rightarrow 1$. However, if one assumes a 
pure hydrodynamic cascade in which the fast magneto-sonic modes are dissipated at weak shocks, the 
cascade cutoff scale becomes very large and both TTD and NR mechanisms become very inefficient. 
This is illustrated in the bottom left panel of Figure~\ref{f7:fig}, where the total advection rate is now  
completely dominated by AdC falls below he min loss rate except during the major merger event.
Finally, in the BL11 scenario (bottom right panel) the TTD advection rate is highly boosted by the 
effective collisionality of the thermal plasma, which suppresses the damping of the fast magneto-sonic
waves responsible for the acceleration of the particles. The plots assume a cosmic-ray energy
fraction $x_{CR}=10\%$, no intermittency, $\xi=100\%$, as well as Kraichnan's cascade model.
The NR mechanism is much less affected for the same reasons as above, and its rates  
comparable to those in the BL7 scenario. Obviously AdC is not affected, as in the previous panels.
As already mentioned, when $\zeta_2\rightarrow 1$ TTD mechanism is suppressed even in the case
of high plasma collisionality.

We now turn to the properties of the emitted radiation in the various scenarios.
While the time evolution of the emitting particles could be integrated in time using a numerical
scheme~\citep[see e.g.][]{2001CoPhC.141...17M,2007JCoPh.227..776M}, 
here we adopt a simplified and more straightforward
analytic method, leaving the more accurate treatment for the future.
In particular we are interested in the evolution of the spectral index 
of the synchrotron radiation of the emitting particles
\begin{gather}\label{alpha:eq}
\alpha\equiv-\frac{\partial\log j_{syn}(\nu)}{\partial \log\nu}.
\end{gather}
We assume conditions of approximate steady-state equilibrium. This apply for example when,
during a merger event, the rise of turbulent energy is about to turn around and the distribution 
function time derivative to change sign. We also consider the multi-GeV energy range, 
where particle cooling rate, $\Gamma_{syn+IC}$, 
is dominated by synchrotron and inverse Compton mechanism~\citep{rpa79}. 
Under these conditions the equilibrium particle distribution function is simply
\begin{gather} \label{f:eq} 
f(p)=f_0e^{-q} , \quad
q\approx \frac{4\Gamma_{syn+IC}}{\Gamma_p},
\end{gather}
and the synchrotron spectrum
\begin{gather}\label{jsy:eq}
j_{syn}(\nu)\propto \left(\frac{2\nu}{3\nu_B}\right)^{\frac{3}{2}}
\int dx \frac{F(x)}{x^{\frac{5}{2}}} \int d\mu \frac{e^{-q(x,\mu)}}{(1-\mu^2)^\frac{1}{4}},
\end{gather}
where $F(x)=x\int_xK_{5/3}(z)dz$, $K_{5/3}$ is the modified Bessel's function of order 5/3 and
$\mu$ is the particles pitch angle, assumed uniformly distributed.
The magnetic field entering the cyclotron frequency is chosen to be $B_{CMB}$ so that
$\nu_B=\nu_{CMB}$, corresponding to the most favourable conditions for
synchrotron emission (see above).
It is then easy to carry out the log-derivative of the emissivity in Eq.~(\ref{jsy:eq})
to find out the time dependent spectral index defined in Eq.~(\ref{alpha:eq}) characterising 
the various scenarios discussed above in connection with Figure~\ref{f7:fig}. 
This involves some integration which is straightforwardly carried out numerically.

Note that when $\Gamma_p<\Gamma_{syn+IC}$, the particles are not advected towards higher 
values in momentum space but can still be accelerated diffusively~\citep{2011MNRAS.410..127B}.
This is in fact what produces the exponential roll-off in the distribution function of Eq.~\ref{f:eq}. 
And it is these particles in the exponential tail that are thought to produce the steepening spectrum 
characteristic of the diffuse radio emission observed in galaxy clusters.

The results are presented in Figure~\ref{f8:fig}, for emission at 300 MHz (solid line), 1.4 GHz (dotted line)
and 8 GHz (dashed line). The top two panels and the bottom left panel correspond, respectively,
to the BL7, BL7k and BL7b scenarios (as the corresponding panels of Figure~\ref{f7:fig}),
while the bottom panel is for the BL11 scenario assuming $x_{CR}=10\%$ and $\xi=10\%$ (vs
 $x_{CR}=10\%$ and $\xi=100\%$ in Figure~\ref{f7:fig}).
The time span is restricted to the few Gyr around the time of high mass accretion rate. 
The plots show that in general the time dependence of the spectral index is, as expected from
the previous figure, flattening and steepening before and after core passage, respectively,
with the minimum value corresponding to the peak of the advection rate during the major merger. 
In the BL7 case the spectral index at 300 MHz is between 0 and 1.5 for about a Gyr , between
0.2 and 2 for almost a Gyr at 1.4GHz, and between 2 and 4 at 8GHz. These values are reasonable 
for classical radio halos with spectral index around 1~\citep{2008SSRv..134...93F,2012A&ARv..20...54F}.
Steeper spectral indexes are predicted in the slightly less efficient cases of BL7k, where
$\alpha>0.5-2$ and $>1.5-4$, at 300 MHz, 1.4 GHz, respectively, only partially viable for
classical radio halos but suitable for the 
ultra-steep halos~\citep[$\alpha\gtrsim 2$,][]{2010A&A...509A..68C,2008Natur.455..944B,2013ApJ...777..141C},
while in the inefficient BL7b case the spectral index is very large 
and basically inconsistent with the observations.
Note that according to Eq.~(\ref{Gamin:eq}) the energy losses grow
quite rapidly with redshift, as $(1+z)^\frac{7}{2}$. So the situation
may become also challenging for the BL7 scenario at higher redshifts, 
namely $z\simeq 0.4--0.5$. However, this topic deserved a separate and 
more focused discussion and investigation, based on the specific 
characteristics of halos at those redshifts.
In the case of BL11 the advection rates are much higher,
so one expects lower spectral indexes. This is indeed the case.
When $x_{CR}=10\%$ and $\xi=100\%$, corresponding to the bottom-right panel of
Figure~\ref{f7:fig} we obtain results inconsistent with observations.
We therefore consider the case $x_{CR}=10\%$ and $\xi=10\%$. For these parameters
the spectra during the major merger are still rather flat. At low frequencies
they remain within the observed range even prior the major merging, i.e. 
during the phase of enhanced mass accretion including the (triple) minor merger, 
so the sources acquires a long lifetime of almost 3 Gyr (the lifetime being 
roughly defined as the time interval during which the spectral index 
remains consistent with observed values).

\section{discussion}  \label{t_disc:sec}

The analysis in the preceding Section indicates that the acceleration
rate due to stochastic acceleration processes increase significantly
during the episode of high mass accretion rate, peaking in value
during the major merger event.  This is in general, qualitative
agreement with requirements from astronomical observations, indicating
that sources of diffuse radio emission in GC occur in connection with
energetic mergers. Important details related to the microphysics of
the ICM remain to be understood, however, as they affect significantly the
quantitative predictions of the model. We discuss these briefly in the following.

In the BL7 scenario, which assumes
Kraichnan's cascade for fast magnetosonic waves, the acceleration rate
appears sufficient to power synchrotron spectra with consistent with
classical radio halos.  When the assumption of Kraichnan's model are
relaxed towards the steeper cascade model produced by the simulation
and based on partial dissipation by weak shocks, the situation worsens
noticeably. In this case it becomes difficult to reconcile the results
with observations.  Thus the nature of the cascade of fast magneto
sonic waves in the ICM has a fundamental impact on the acceleration
efficiency of both TTD and NR mechanisms.  It would therefore be very
valuable to understand to what extent Kraichnan's scenario applies and
to what extent dissipation by weak shocks steepens the cascade.
Finally, in the plain version of the BL11 scenario, the acceleration
efficiency become potentially too high. Here the turbulence is mainly
dissipated via TTD by cosmic-ray protons in the ICM and the diffusion
coefficients is inversely proportional to their energy density (see
Eq.~\ref{dpp_c:eq}).  This quantity has been constrained by several
gamma-ray experiments to $x_{CR}\le
1\%$~\citep{2014ApJ...787...18A,2013A&A...560A..64H,2012A&A...541A..99A}.
In our estimate we have relaxed this upper limit to
10\%~\cite[assuming, e.g., the cosmic-ray spatial distribution is
broader than the thermal gas,][]{2001ApJ...559...59M}.  Of course in
this case, secondary $e^+e^-$ generated in hadronic collisions by the
above cosmic-ray protons might also produce significant radio emission
in the magnetic field given by
Eq.~(\ref{Bmin:eq})~\cite[e.g.][]{2003MNRAS.342.1009M}.  In addition
we have assumed a reduced effect of micro-instabilities on the plasma
collisionality such that $\xi\approx 10\%$. In this case we obtain
spectra that perhaps remains too flat, but in a reasonable range, and
a source lifetime roughly consistent with observations. Of course the
above parameters such as $x_{CR}$ and $\xi$ ideally should be
determined independently and {\it not} chosen to reproduce the
observations.  In any case, the microphysics apparently plays a
crucial role here. The important implications of cosmic-ray feedback
was also discussed by~\cite{2013ApJ...771..131B,2014IJMPD..2330007B}.
It is important to remember that we have so far considered a specific magnetic field, $B_{min}$
(see Eq.~\ref{Bmin:eq}). Without entering a full discussion on this quite complex topic,
we should at least mention that larger ICM magnetic fields than $B_{min}$
are likely and would favour the collisional scenario.

It is worth noting that while the radiative losses at fixed frequency increase rapidly with 
redshift (see Eq.~\ref{Gamin:eq}), RH are observed up to $z=0.5$, with a constant number 
of sources per redshift interval out to $z=0.3$. These objects are typically more massive than
the object in our simulation, so their turbulent motions and predicted acceleration rates will be 
correspondingly higher. 
But if we extrapolate our simulated cluster ($M<2\times 10^{15}M_\odot$) at these modest $z$ 
we find the acceleration rates in the BL7 scenario perhaps insufficient for RHs.
Thus a discovery of a population Coma-like RHs at these and higher redshift will rule out 
purely collision less flavour of compressional turbulence acceleration in the ICM.
This might soon be checked by surveys with LOFAR and more in the future with SKA.
It is worth reminding that the major merger experienced by the simulated GC is a rarely 
strong one (Section~\ref{ah:sec}), with a relative velocity occurring in only 5\% of the cases. 
Thus the acceleration rates are not limited by of a particularly weak merger. 

In addition, the pdf-statistics of the increments of compressional
velocity in Figure~\ref{f4:fig} shows that very high velocity
increments above 10$^3$ km s$^{-1}$ occupy less than 10\% of the
volume. Sufficiently strong acceleration might occur in a sub-volume
in which the turbulence is particularly strong.  However, in order for
this sub-volume would to have covering factor of order unity, the
emitting regions would have to be rather thin (for example, for a 10\%
subvolume implied by the above statistics, arranged to cover a surface
of diameter 1 Mpc, one infer a thickness less than 0.1 Mpc). This
might lead to polarised emission, a possibility which should be
investigated in more details in the future.

Finally, from the results in the previous section it is clear that
sufficient compressional turbulence is generated only close to core
passage of the major merger event. The origin of the classical,
moderately steep radio halo of the Coma cluster, therefore, remains a
puzzle in the less efficient scenarios, given that this cluster
experienced his the last major merger about 2 Gyr
ago~\citep{1994ApJ...427L..87B}.  In addition, while BL11 case with
small $x_{CR}$ and large $\xi$ is sufficiently efficient to produce
diffuse radio emission even during minor merger events, such as the
one possibly currently experienced by Coma cluster, it is essential to
understand how those parameters change with changing turbulent
conditions in the plasma (constant values would be inconsistent with
observations).

So, perhaps the most important conclusion emerging from the above
discussion is that RH are complex phenomena in which a hierarchy of
processes are at work. Unlike other cases in astrophysics the
subtleties of the ICM plasma microphysics here appear to play an
important role with a potentially strong impact.  The study of RH
therefore offers a unique opportunity to learn about these processes.
In this context the ability to model reliably the ICM turbulence with
high resolution simulations is instrumental because it removes an
important unknown of the problem, namely the turbulence statistics,
and allows to start addressing these more elusive processes that
characterise the acceleration physics.  Obviously, while in this paper
we have used the properties of a single GC, it will be important to
improve on the analysis, extend it to a representative sample of
simulated GC and explore the fully MHD case, amongst others.
Eventually, these efforts might be corroborated by measurements of the
ICM turbulence thanks to the advent of high precision X-ray
spectrometers such as Astro-H\footnote{http://astro-h.isas.jaxa.jp}
and Athena\footnote{http://sci.esa.int/ixo}, which will have energy
resolution of a few eV, and perhaps the help of innovative analysis
techniques~\citep{2004A&A...426..387S,2004MNRAS.347...29C,2011MNRAS.410.1797S,2012MNRAS.422.2712Z,2013A&A...559A..78G,2014ApJ...788L..13Z,2014A&A...569A..67G}.

Another important point to constrain the physical scenario is to 
continue verifying the connection between the compressional turbulence driving the 
stochastic acceleration of the supra-thermal particles, and the RH 
properties~\citep[e.g.][]{2010ApJ...721L..82C,2011MNRAS.417L...1R,2012MNRAS.421L.112B,2013AN....334..350B,2013ApJ...777..141C}. 
This can be achieved by looking at other testable implications emerging from this model.
There are two such potential tests. Although they might prove not
straightforward, they may become feasible with future surveys.
First, we note that in Figure~\ref{f7:fig}
the peaks in the total acceleration rate correspond to core passage during the two 
mergers events, while the roughly linear rise and fall around the peaks is correlated with the 
nearing and departing of the merging structures. This is of course due to gas compression and
decompression following the passage of the substructure through the main cluster.
Thus strong compressional turbulence persists 
during $\tau_{c}\approx 2d/v_{rel}$, where $d\approx L\approx R_{vir}/3$ is the distance at which 
turbulence injection begins, and $v_{rel}\approx a_v v_{vir}$, the relative velocity of the merging 
structures with $a_v\gtrsim 1,$ and $R_{vir}$ and $v_{vir}$, the virial radius and velocity 
respectively. Thus the simulation also predicts a correlation of radio power with distance 
of the merging cores.  In addition, as discussed in Section~\ref{adr:sec}
and shown in Figure~\ref{f8:fig}, the spectral index is time dependent and should get flatter with the
approaching of the merging cores, and steeper with the departing of the same. Thus, like the radiation
power, the spectral index should show some correlation with distance of the merging cores.
Based on the level of compressional turbulence alone,
these correlations should be symmetric with respect to time since core passage.
Actually in both cases there will be a 'space-lag' corresponding to the distance traveled by the merging 
cores during a cooling time, i.e. $d_{lag}\sim v_{rel}\Gamma_p^{-1}$.
In addition, more subtle complications may arise depending on the detail of the acting microphysics.
In any case some correlation should emerge above the noise, particularly for strong merger 
events, provided a sufficiently large number of RH sources becomes available. 
While the statistical details are important, they are not attempted here as beyond the scope of the paper.

\section{Summary and Conclusions} \label{sum:sec}

In this paper we have used results from the Matryoshka run~\citep{2014ApJ...782...21M} to 
model the time dependent statistical properties of the turbulence in the ICM of a massive GC. 
The very high resolution of the simulation across the virial volume allows us to model the 
cascade of hydrodynamic turbulence and capture reliably its inertial range and outer scale. 
We have also decomposed the velocity field into compressional and incompressible components
and measured for each case the longitudinal and transverse structure functions, particularly those of 
second order which are of interest in this study. 
We have looked at the history of turbulent energy and its fraction in compressional motions,
and made a connection to the mass accretion history of the GC. One interesting point is that 
at late time, $z\approx0$ the GC experiences a major merger, raising significantly the level turbulent 
energy and the compressional fraction thereof. However, the mass accretion rate and, consequently,
the turbulent energy increase significantly for an extended period of time which is longer than 
the duration of the merger. We have ascribed this to the fact that spatial clustering of 
mass gives rise to a temporal clustering of mass accretion~\citep{Peebles93}.
We have also looked at other statistical properties of the turbulence, in particular the 
ratio of the second order transverse and longitudinal structure functions. This reveals 
that the conditions of the flow are often approximately consistent with fully developed 
turbulence, as departures following merger events settle in about an eddy turnover time.
In terms of velocity statistics we have considered the pdf of the velocity increments at the 
injection scale. We have found that while the median value is negligible for the incompressible 
component, it is negative for the compressional velocity, indicating a systematic convergence 
of weak shocks and adiabatic heating of the particles scattering off them.

We have extracted the relevant turbulence properties to estimate the particle acceleration rates
due to stochastic processes, and compared their predictions with observations of diffuse radio 
emission in GC.
In particular, we have considered  the TTD~\citep{1976JGR....81.4633F,1991ApJ...376..342M} and NR~\citep{1988SvAL...14..255P}
mechanism, considered in connection with the origin of RH phenomena in BL7 and BL11.
These mechanisms are based on the interactions of the particles with compressional
waves which, in the high-$\beta$ ICM plasma, are represented primarily by 
fast magneto-sonic waves. 
The acceleration rates depend on both the amount of turbulent energy but also 
on detailed aspects of the physics of the ICM plasma determining the cascade of the 
compressional waves. These aspects are not captured by the simulation.
However, a dependable numerical models such as this,
removes from the problem the unknown represented by the hydrodynamical turbulence,
allowing the investigation of assumptions concerning micro-physics of the ICM plasma.
The results reported in this paper indicate that the microphysics of the ICM plasma 
affects significantly the efficiency of the above stochastic acceleration processes.
Particularly important are the details of the cascade of  fast magneto-sonic waves
affected by the plasma collisionality of the thermal ICM and the effectiveness of dissipation
processes such as weak shocks. The TTD mechanism can become very efficient,
even during minor mergers, but this requires understanding of how 
plasma microscopic behaviour changes during the such mergers, 
as the predictions become otherwise inconsistent with observations.
We have proposed two new tests to collect additional evidence for the role of stochastic processes in RH~\citep[e.g.][]{2010ApJ...721L..82C,2011MNRAS.417L...1R,2012MNRAS.421L.112B,2013AN....334..350B,2013ApJ...777..141C} based
on the rapid rise and fall of the population of emitting relativistic particles, following the rise and fall
of the turbulence. The tests basically concern a statistical correlation 
of the radio power and the spectral index with distance of the merging cores.
Given the statistical nature of these tests, large catalogues of RH are probably necessary
which could become available with new surveys by powerful instruments such as LOFAR and SKA.
\\

\acknowledgements 
The author gratefully acknowledges several useful discussions with G. Brunetti, not least
during a visit at the hospitable Istituto di Radio Astronomia. This work was supported by a grant from the
Swiss National Supercomputing Center (CSCS) under project ID S275 and S419.

\bibliographystyle{apj}
\bibliography{papers,refs}

\end{document}